\documentclass[10pt,twocolumn, aps, prb, superscriptaddress]{revtex4-1}
\usepackage{graphicx}
\usepackage{amsmath}
\usepackage{amsfonts}
\usepackage{amssymb}
\usepackage{tabularx}
\usepackage{color}
\newcommand{\kvec}{\boldsymbol{k}}
\newcommand{\qvec}{\boldsymbol{q}}
\newcommand{\lvec}{\boldsymbol{l}} 
\newcommand{\Kvec}{\boldsymbol{K}}
\newcommand{\nvec}{\boldsymbol{n}}
\newcommand{\mvec}{\boldsymbol{m}}
\newcommand{\rvec}{\boldsymbol{r}}

\begin{document}

\begin{abstract}
We use numerically exact quantum Monte Carlo (QMC) to compute the
properties of three dimensional bipolarons for interaction strengths
where perturbation theory fails. For intermediate electron-phonon
coupling and Hubbard $U$, we find that bipolarons can be both small
and light, a prerequisite for bipolaron superconductivity. We use the
QMC results to make estimates of transition temperatures,
which peak at between $90-120$ K and are demonstrated to be insensitive to
Coulomb repulsion and impurities.
\end{abstract}

\title{Mobile Small Bipolarons on a Three-Dimensional Cubic Lattice}

\author{A. R. Davenport}
\affiliation{Department of Physical Sciences, The Open University, Walton Hall, Milton Keynes MK7 6AA, UK}

\author{J. P. Hague}
\affiliation{Department of Physical Sciences, The Open University, Walton Hall, Milton Keynes MK7 6AA, UK}

\author{P. E. Kornilovitch}
\affiliation{Hewlett-Packard Company, 1070 NE Circle Boulevard, Corvallis, Oregon 97330, USA}

\maketitle

\section{Introduction}
\label{polandBipol}

The mechanisms of three-dimensional superconductors with high
transition temperatures (such as the bismuthates) are widely
acknowledged to have their origins in the electron-phonon
interaction. In at least some materials, a proposed mechanism involves
the pairing of phonon dressed electrons (polarons) to form
bipolarons. At sufficiently low temperatures, these bipolarons form a
Bose--Einstein condensate with superconducting
properties\cite{Alexandrov1981a}. The main criteria for bipolaronic
superconductivity with significant transition temperatures are
bipolaron mobility (low effective mass) \cite{Edmin1989} with
sufficient density (small bipolarons). Although small (bi)polarons are
often viewed as immobile states that may be easily localized by
disorder, small mobile bipolarons that can lead to superconductivity
with high transition temperatures have been found in 1 and 2
dimensions. \cite{Edmin1989,PolaronMaterials,hague2007a}

Polarons occur naturally in nearly all media, from plasmas and ultra
cold atoms through normal bulk materials and possibly high-temperature
superconductors. \cite{PolaronObservation, PolaronMaterials}
Landau introduced the polaron concept in 1933 to
study lattice polarization due to the motion of electrons through
 ionic solids \cite{PolaronProblem}: An electron moving though a crystal
lattice creates distortions that follow its trajectory, producing a
phonon cloud that propagates though the system, surrounding the electron.
\cite{PolaronMaterials, hague2009a}

Bipolarons are formed when two polarons interact with each
other using phonon-mediated interactions to form pairs. These pairs can be strongly bound and travel
though the lattice as a single composite particle. There are many
different interaction types that lead to bipolaron creation.
\cite{hague2010a} In this paper, we consider an extended
Hubbard--Holstein model, where the Coulomb interaction is purely
on-site, and an extended-Holstein interaction couples the electron
density to lattice vibrations on the same site and near-neighbor sites,
similar to the interaction introduced by Bon\v{c}a and Trugman.
\cite{bonca2001a,hague2010a} Increasing this inter-site interaction allows
electrons to form bipolarons and overcomes the Coulomb repulsion so
that large bipolarons become local (small) bipolarons with
approximately the size of the lattice constant.
\cite{Alexandrov2002a} The site-local Holstein model describes an
extreme short-range limit where electrons can form pairs on single
atoms. In this case, the electron-phonon interaction has to overcome
the on-site repulsion to form a bipolaron. \cite{Wellein1996a} We
note that the interactions used here are distinct from the
specific forms required to describe bismuthate superconductors. \cite{bischofs2002a}

Recent studies of low-dimensional bipolarons have utilized various numerical 
methods and analytical techniques. One-dimensional bipolarons
have been found to be relevant in describing strong electron-phonon 
interactions in low-dimensional organic semiconductors, \cite{OrganicBipolaron} 
and it is possible that three-dimensional (3D) bipolarons in a strong magnetic field simplify
into one-dimensional (1D) bipolarons. \cite{3DBipol+Field=1DBipol}
Two-dimensional (2D) bipolarons have been investigated extensively in the study of two 
dimensional conductors and high temperature superconductors such 
as the cuprates. \cite{Alexandrov1981a}

Properties of the short range Hubbard-Holstein bipolaron model have been established 
on small lattices using exact diagonalization \cite{Wellein1996a} 
and an optimized approach for exact diagonalization at weak 
coupling. \cite{Weibe2000a} Advanced variational techniques, \cite{ Bonca2000a}
density-matrix renormalization-group, \cite{jackelmann1999a}
and various quantum Monte Catlo (QMC) approaches \cite{hague2009a, Hohenadler2005a} 
have all been used to study 2D bipolaron systems. 
It is found in one dimension that only a small attractive force between 
electrons leads to pairing, the most important factor being the nearest 
neighbor interaction when considering long ranged interactions. \cite{hague2009a,1DExtHolstein}
Two dimensional work also shows us that inclusion of nearest-neighbor interaction is responsible 
for significant change compared with the Hubbard-Holstein model. \cite{2DHubbard, hague2010a}
Light bipolarons are found on a simple square lattice, showing that elaborate lattices 
are not needed to create small light pairs that have the potential to form Bose 
condensates. \cite{hague2010a} On change of dimension from 1D to 2D, an increase 
in electron-phonon coupling constant and nearest neighbor attraction is needed to create on-site bipolarons. \cite{hague2009a,hague2010a}

There have been several notable studies of 3D bipolarons, especially
in relation to superconductors, to understand the reasons why cuprates
and other layered superconductors are different to 3D materials.
\cite{Sil1991a,Devreese1991a} Several publications have examined the
differences between the binding of bipolarons in two and three
dimensions, concluding that a greater attraction is needed to form
stable bipolarons on 3D lattices.
\cite{GroundStateBipolaons,BipolVariationalStrongCoupling}  It has
been found that the probability of bipolaron formation increases with
decreasing dimensions or increase in the crystal anisotropy.
\cite{Kashirina2010a}  Variational studies of the region of existence
of the three-dimensional singlet bipolaron have allowed investigation
of the relationship between the critical value of the electron-phonon
coupling constant and the dielectric properties of the medium,
\cite{Mukhomorov2004a} concluding that conditions in alkali-halide
crystals are not suitable even for metastable bipolarons and that
three-dimensional continuum bipolarons do not exist in La$_2$CuO$_4$.
However, metal-ammonia systems potentially lie within the region of
existence for three-dimensional continuum bipolarons.
\cite{Mukhomorov2004a} In the context of 3D polarons, we have also
examined binding to attractive impurities, showing that polarons are
localized when the impurity potential is around four effective hoppings
in magnitude. \cite{hague2008a}

The work presented here goes beyond previous work by considering exact
solutions for 3D extended Hubbard--Holstein bipolarons (both numerical
and analytic). Exact solutions are important to understand regions of
the parameter space where perturbative approximations break down, and
as we will show are essential for understanding the regions where
bipolaronic superconductivity is strongest. The paper is organized as
follows: In Sec. \ref{sec:modelmethods}, we introduce the model. The
Lang-Firsov transformation is performed before a brief overview of the
continuous time quantum Monte Carlo simulation method. Sec.
\ref{sec:MonteCarlo} presents quantum Monte Carlo results for singlet
bipolaron properties, including total energy, number of associated
phonons, inverse mass, and average bipolaron size. Finally in Sec
\ref{Sec:super}, we look at the possibility of Bose Einstein
condensation of three-dimensional bipolarons. For completeness, in the
Appendix, we consider the $U - V$ model
corresponding to the high phonon-frequency limit.  Solving the
equation analytically we find the binding conditions for varying
on-site attraction $U$ and nearest neighbor repulsion $V$.

\section{Model and methods}
\label{sec:modelmethods}
\subsection{Model}
The extended Hubbard--Holstein model used here has its basis in a
general electron-phonon Hamiltonian with electrostatic repulsion,
which is written in the following form \cite{hague2010a, hague2009a}:
\begin{eqnarray}
H &=& \displaystyle-t\sum_{<\mathbf{nn}^\prime>,\sigma} c^\dagger_{\nvec^\prime,\sigma}
c_{\nvec,\sigma} \nonumber \\
&+& \frac{1}{2} \sum_{\nvec\nvec^\prime\sigma\sigma^\prime} v(\nvec,\nvec^\prime) 
c_{\nvec\sigma}^{\dagger} c_{\nvec\sigma} c^\dagger_{\nvec^\prime\sigma^\prime} 
c_{\nvec^\prime\sigma^\prime} + \sum_{\mvec} \frac{\hat{P}^2_{\mvec}}{2M} \nonumber \\
&+& \sum_{\mvec} \frac{\xi^2_{\mvec} M\omega^2}{2} - \sum_{\nvec\mvec\sigma} f_{\mvec}(\nvec)c^\dagger_{\nvec\sigma}c_{\nvec\sigma}\xi_{\mvec}
\label{eqn:HubbardFroehlich}
\end{eqnarray}
where $\nvec$ and $\mvec$ represent vectors to electrons and
ions respectively, $c \left( c^\dagger\right)$ are the creation
(annihilation) operators for electrons, $M$ is the ion mass, $\omega$ is the phonon
frequency and $\sigma$ is the $z$ component of the electron spin. The
first term in the equation expresses the kinetic energy of electrons
moving from site to site. The element $t$ is the hopping integral
for an electron moving between neighboring sites.

The second term in the Hamiltonian represents the Coulomb repulsion
$v$ between two electrons. Here the repulsion is approximated to have
the Hubbard form, and long ranged interactions are assumed to be
insignificant due to screening in the material \cite{hubbard1963a} so
the repulsive term has the form $H_{\text{Hubbard}} = U \sum_n n_{i
\uparrow} n_{i \downarrow}$, where $U$ is the magnitude of the
repulsion \cite{hague2010a, hague2009a}. Note that
near-neighbor interactions are in principle allowed, but neglected in 
section \ref{sec:MonteCarlo} and appendix \ref{sec:HighFrequency} of this paper.

The final three terms include the effects of lattice vibration. The
ion momentum is described by the $\hat{P}_{\mvec}$ operator and the
ion displacement is signified by $\xi_{\mvec}$. Here we take
$\xi_{\mvec}$ to be one dimensional, which is an approximation that
could relate to phonon modes polarized by a strong electric field, a
three-dimensional molecular crystal with molecular ordering along a
single direction or possibly radial phonon modes.  Following
Ref. \onlinecite{hague2010a}, we take the force function to be,
\begin{eqnarray}
f_{\mvec}(\nvec) = \kappa \sum_{\lvec_{i}}\delta_{\nvec,\mvec+\lvec_{i}/2}
\label{eqn:forcingfunction}
\end{eqnarray}
This describes interaction between electrons on sites at vectors
$\nvec$ and vibrating ions between valance sites at positions
$\mvec$. $\lvec_i$ are the vectors between nearest neighbor valence
sites at $\rvec$ and $\rvec^\prime$. An effective electron-electron interaction can be defined as
\begin{gather}
\Phi_{\Delta \rvec}[\rvec,\rvec^\prime] = \sum_{\text{m}}
f_{\text{m}}[\rvec] f_{\text{m}+\Delta\rvec}[\rvec^\prime] \label{eqn:CompPhi}
\end{gather}
so that for the chosen force function, $\Phi_{0}[\rvec,\rvec^\prime]/\Phi_{0}[0,0]=\gamma$
where the nearest neighbor interaction strength ($\gamma$) strictly has the value
$1/z$. The reason for $\Delta \rvec$ will be explained later on in the paper. 
Here we will also modify $\gamma$ to investigate the effects of
turning on the inter-site interaction. For large phonon frequency,
this interaction can then be mapped directly onto a $U-V$ model. For
$\gamma=0$, a Holstein interaction is recovered, equivalent to $
f_{\mvec}(\nvec) =\kappa\delta_{\mvec\nvec} $. The shift in $\gamma$ is 
equivalent to moving the vibrating ions within the unit cell so that they 
get closer to the sites host electrons.

\subsection{Lang-Firsov transformation}
In the limit that phonon frequency becomes infinite; the model described in Eq. (\ref{eqn:HubbardFroehlich}) can be
mapped onto a $U-V$ model, consisting of an on-site Hubbard $U$ and
inter-site Hubbard $V$. The mapping uses a Lang-Firsov canonical transformation
\cite{langfirsov}, which creates a new Hamiltonian $\tilde{H} = e^{-S}He^S$
and wavefunction $\tilde{|\psi\rangle} =
e^{-S}|\psi\rangle$, where $\tilde{H}= H + [S,H] + [S,[S,H]] + \cdots$
and $S = \text{g}n(d^\dagger - d)$. Here g is a dimensionless
interaction constant proportional to the force and $d^\dagger(d)$ is
the phonon creation (annihilation) operator. Under this
transformation, the creation operators for electrons and phonons
become,
\begin{equation}\centering\begin{split}
c^{\dagger}_i \rightarrow \tilde{c}^\dagger_i &= c^\dagger_i \text{exp} 
\left[ \sum_j \text{g}_{ij} \left(d^\dagger_j - d_j\right)\right] \\
d^\dagger_j &\rightarrow \tilde{d}^\dagger_j = d^\dagger_j + \sum_i \text{g}_{ij}n_i \label{eqn:LFT1}
\end{split}\end{equation}
On transforming the atomic Hamiltonian ($t\rightarrow 0$) the electron and phonon subsystems are decoupled:
\begin{equation}
\tilde{H}_{at} = - \sum_{ii^\prime}n_in_{i^\prime}\sum_j \frac{f_{ij}f_{i^\prime j}}{2M\omega^2} + 
\hbar\omega \sum_j \left(d^\dagger_jd_j + \frac{1}{2}\right),
\end{equation}
The function $\Phi_{0}$ and a
dimensionless interaction parameter $\lambda = E_p/W$ are introduced
to simplify the Hamiltonian, where $W$ is the half band-width $zt$,
and $E_{p} = \sum_j f^2_{0j}/2M\omega^2 = \Phi_0(0,0)/2M\omega^2$ is the polaron shift, leading to:
\begin{gather}
\tilde{H}_{at} = - \sum_{ii^\prime}n_in_{i^\prime}\frac{W\lambda\Phi_0(i,i^\prime)}{\Phi_0(0,0)} + 
\hbar\omega \sum_j \left(d^\dagger_jd_j + \frac{1}{2}\right) \label{eqn:ATratio}.
\end{gather}
Transformation of the tight-binding Hamiltonian leads to
\begin{gather}
\tilde{H}_{tb} = \sum_{ii^\prime}\sigma_{ii^\prime}c^\dagger_ic_{i^\prime},
\end{gather}
with,
\begin{multline}
\sigma_{ii^\prime} = t_{ii^\prime} \text{exp} \left[ -\frac{W\lambda}{\hbar\omega} \left(1-\frac{\Phi_0(i,i^\prime)}{\Phi_0(0,0)}\right) 
\right] \\ \times \text{exp} \left( \sum_j(\text{g}_{ij}-\text{g}_{i^\prime j})d^\dagger_j \right) \text{exp} 
\left( -\sum_j(\text{g}_{ij}-\text{g}_{i^\prime j})d_j \right)
\end{multline}
Where $\text{g}_{ij} = f_{ij} / \omega\sqrt{2M\omega}$. When the
phonon frequency is very large, the ground state contains no real
phonons, and there is a further simplification leading to a modified
hopping, $\sigma_{ii^\prime} \approx t'_{ii^\prime}=t_{ii^\prime}
\text{exp} \left[ -\frac{W\lambda}{\hbar\omega}
  \left(1-\frac{\Phi_0(i,i^\prime)}{\Phi_0(0,0)}\right)
  \right]$. Exact solutions of the transformed Hamiltonian in the
large phonon frequency limit can be found in the Appendix for
comparison with the numerical results.

\subsection{Computational methods}
\label{PhiDelta}
We use the continuous-time quantum Monte Carlo method, which has been
used to simulate the screened Hubbard-Fr\"ohlich bipolaron in 1D and
2D \cite{Kornilovitch1998a,hague2009a}. A more in-depth
overview of our algorithm has been presented in a previous paper and
so will not be repeated \cite{hague2010a}. The continuous-time
quantum Monte Carlo algorithm is based on path integrals, where each
path ${\bf r}_i(\tau)$ exists in imaginary time and represents a single particle in the
system. The algorithm probes path configurations which are each
assigned a weight $\exp(A)$ where:

\small\begin{multline}
A[\rvec(\tau),\rvec(\tau^\prime)] = \frac{z\lambda \bar{\omega}}{2\Phi_0(0,0)} \\ \times \int_0^{\bar{\beta}} 
\int_0^{\bar{\beta}} d\tau d\tau^{\prime} e^{\frac{-\bar{\omega}\bar{\beta}}{2}} \sum_{i j} \Phi_0 [\rvec_i(\tau)], \rvec_j[(\tau^{\prime})] \\
\times (e^{\bar{\omega}(\frac{\bar{\beta}}{2} - |\tau-\tau^\prime|)} + e^{-\bar{\omega}(\frac{\bar{\beta}}{2} - |\tau-\tau^\prime|)} ) \\
+ \frac{z\lambda \bar{\omega}}{\Phi_0(0,0)} \int_0^{\bar{\beta}} \int_0^{\bar{\beta}} d\tau d\tau^{\prime} e^{-\bar{\omega}\tau} e^{-\bar{\omega}(\bar{\beta}-\bar{\tau})} \\
\times \sum_{i j} ( \Phi_{\Delta \rvec}[\rvec_i(\tau), \rvec_j(\tau^\prime)] - \Phi_0[\rvec_i(\tau), \rvec_j(\tau^\prime)]) \\
- \frac{1}{2} \int_0^\beta v(\rvec_1(\tau), \rvec_2(\tau^\prime)) d\tau,
\end{multline}

\normalsize Here $\Delta \rvec = \rvec(\beta) - \rvec(0)$ is the distance between
the end points of the paths in the non-exchange configuration, the
phonon frequency $\bar{\omega}= \hbar\omega/t$, and inverse
temperature $\bar{\beta}=t/k_BT$. $i=1,2$ and $j=1,2$ represent the
fermion paths. $v(\rvec_1,\rvec_2)=U\delta_{\rvec_1,\rvec_2}$ is an instantaneous
Hubbard repulsion between electrons. The paths lie between the
range $\tau\in{0,\beta}$, and are formed from straight segments
punctuated with `kinks' representing site to site hopping. The
algorithm is used to compute (bi)polaron energy, the number of phonons
in the system, the effective mass of the (bi)polaron and the radius of
the bipolaron.

\section{Quantum Monte-Carlo and intermediate phonon frequency}
\label{sec:MonteCarlo}
Using a continuous-time Quantum Monte Carlo code we simulated the
extended Hubbard-Holstein model with nearest neighbor interaction
strengths of $\gamma = 0,0.25$ and $0.5$ on a cubic lattice. The
magnitude of the nearest-neighbor component of the electron-phonon
interaction has been shown in both one and two dimensions to be the
most significant contributing factor to bipolaron properties
\cite{hague2009a,hague2010a}. The simulation produces exact numerical
solutions for the total energy, average number of excited phonons,
mass and size of singlet bipolarons.

In the following we examine only the singlet bipolaron for a range of
$U/t$ and $\lambda$ at inverse temperature $\bar{\beta} = 14$ with
fixed $\hbar\omega/\rm{W} = 1$, where the half band width $\rm{W}=6t$,
which is towards the lower end of the intermediate phonon frequency
limit. Calculations are carried out on an infinite lattice where particles 
are confined to within 50 lattice spacings of each other, 
which is sufficiently large to avoid the majority of finite
size effects. The most sensitive property to finite size effects is
the bipolaron radius, which does not become infinite when the polarons
are not bound into a bipolaron. All errors are determined using
bootstrap re-sampling on the simulation data and are displayed as 3
standard errors.

\begin{figure}[!ht]
\includegraphics[width=0.483\textwidth]{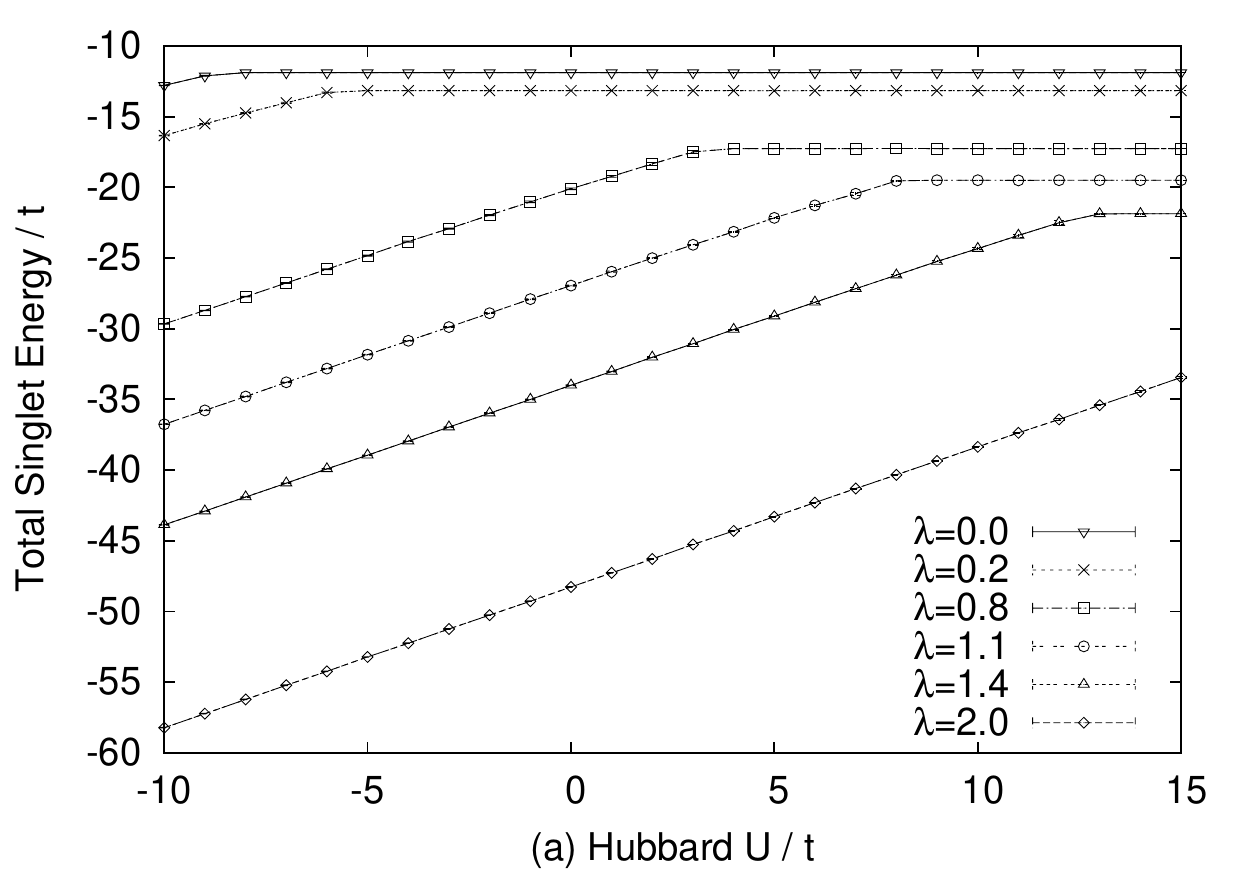}
\includegraphics[width=0.483\textwidth]{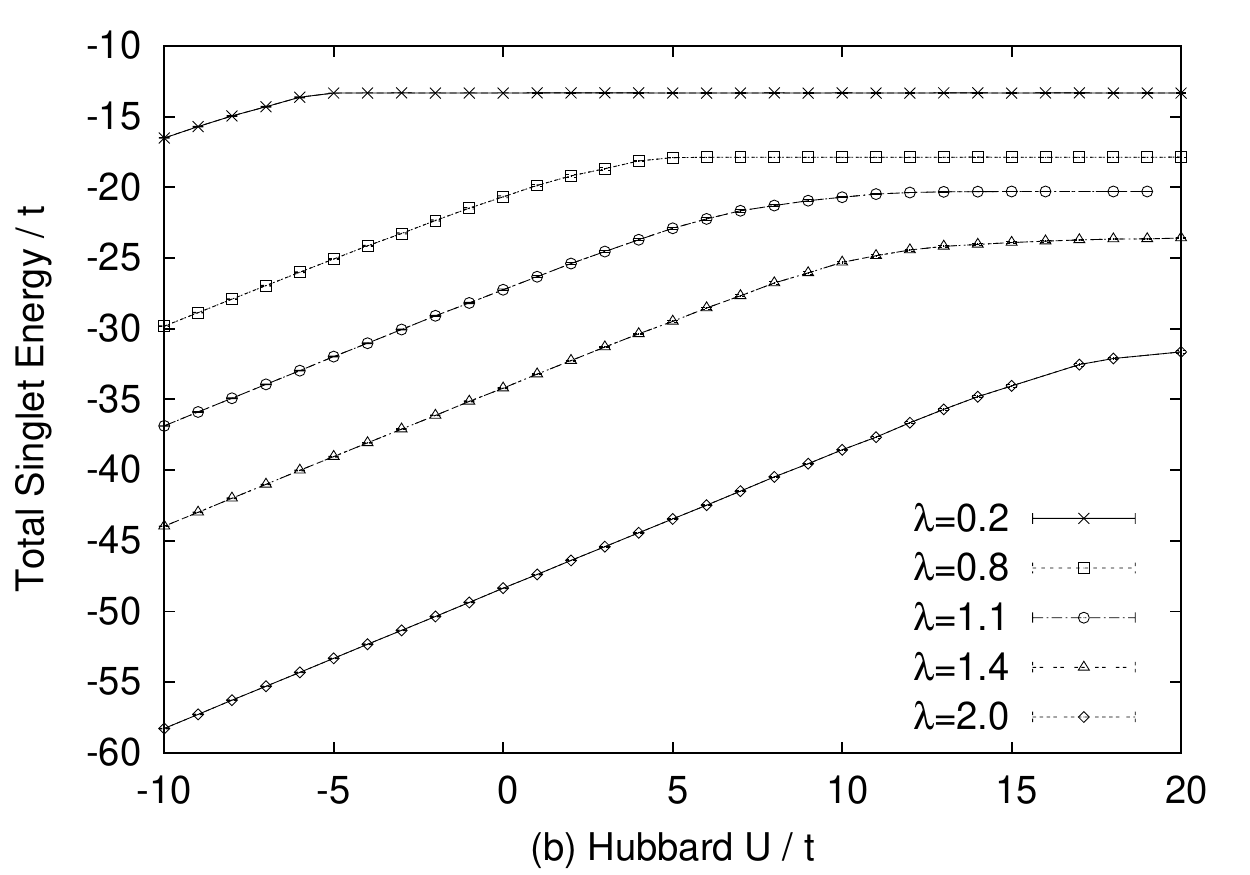}
\includegraphics[width=0.483\textwidth]{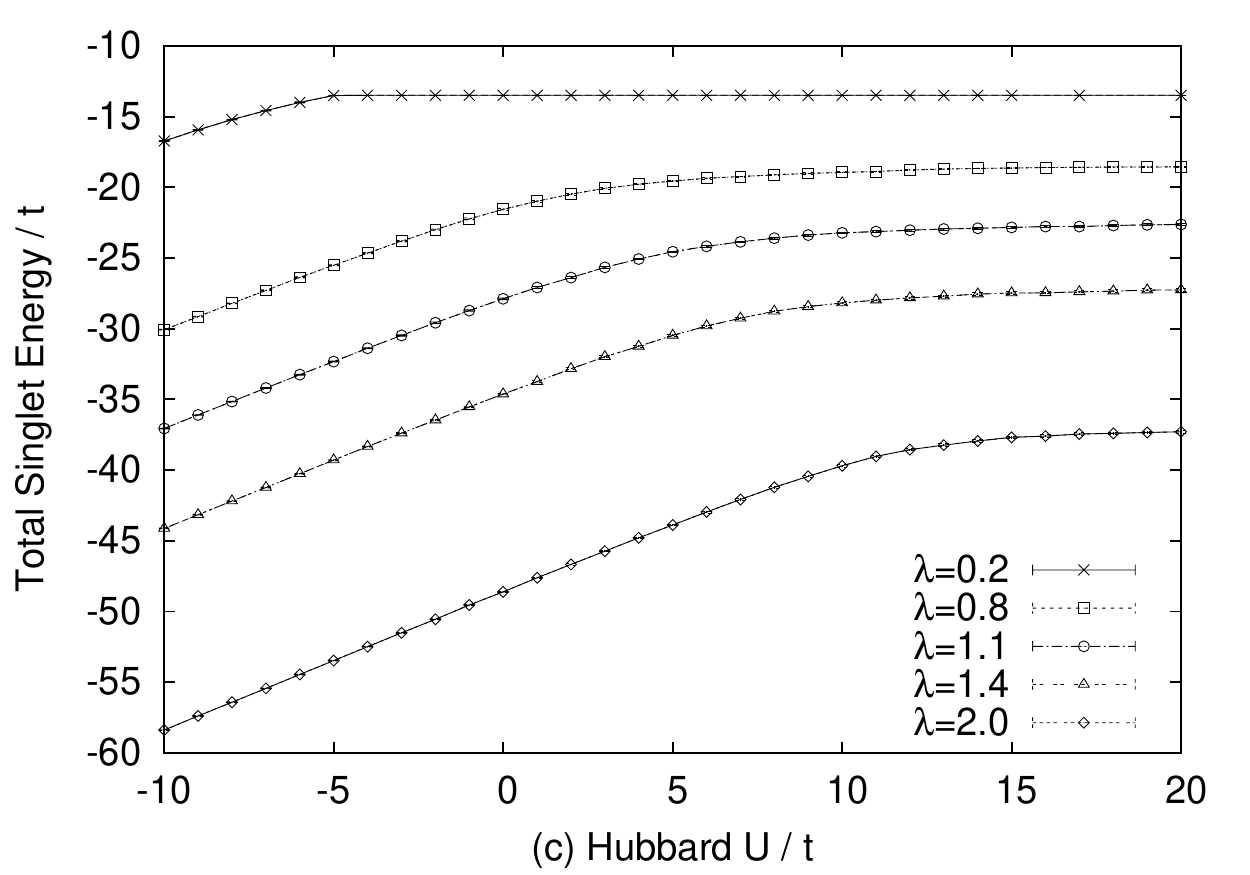}
\caption{Total ground state energy of a singlet Hubbard--Holstein bipolaron
simulated by CTQMC (panel a) and singlet bipolaron with
nearest-neighbor interaction strength of $\gamma = 0.25$ (panel b)
and $\gamma = 0.5$ (panel c). Similar to large phonon frequency, it
takes significant negative Hubbard $U$ to bind on-site
bipolarons. For large inter-site interaction and large $\lambda$, a
crossover is seen between on-site and inter-site
bipolarons.}
\label{Fig:CubicEnergy}
\label{fig:TotalEnergy}
\end{figure}

The binding of bipolarons can be established from the total
energy. Figure \ref{Fig:CubicEnergy}(a) depicts the total energy
calculated from our Monte Carlo calculations when $\gamma=0$ (Holstein
interaction). Diagonal lines show the presence of pairs of electrons
on a single site. As the Hubbard on-site repulsion $U$ is increased,
the on-site pairs are pushed apart creating pairs of polarons. As the
electron-phonon coupling constant $\lambda$ is increased we see that
larger Hubbard $U$ is needed to break apart the on-site pairs. When
the pair is unbound the energy of the bipolaron does not change with
respect to increasing $U$, and the line is horizontal. A key
difference here is that it takes a large negative Hubbard $U$ to bind
on-site bipolarons in contrast to 1D and 2D systems.

Plots of the total energy of the bipolaron formed when the
electron-phonon interaction contains a nearest-neighbor component of
$\gamma = 0.25$ and $\gamma = 0.5$ are also shown 
[Figs. \ref{Fig:CubicEnergy}(b) and \ref{Fig:CubicEnergy}(c)
respectively]. It is immediately apparent that the rapid transition
from on-site bipolaron to free polarons is smoothed out with the
addition of nearest-neighbor interaction. There is no dramatic change
between the total energy of the Holstein and extended Holstein
bipolarons for electron-phonon coupling $\lambda = 0.2$, as there is
insufficient inter-site interaction to bind an off-site bipolaron. This
is in contrast to 1D and 2D where significant qualitative changes to
all bipolaron properties are found when inter-site interaction is
switched on. 

We observe that the $U$ value corresponding to the point of
inflection is reduced with increasing $\gamma$. This does not
correspond to bipolarons which are more weakly bound (i.e. easier to
unbind when the Hubbard $U$ is switched on). Rather, as
electron-phonon coupling is increased, the crossover from bound pairs
to unbound pairs occupies a wider range of $U$ values. Comparison
with Fig. \ref{Fig:CubicEnergy}(a) shows that total energy at large $U$ typically
decreases with increased nearest-neighbor attraction, consistent with
this observation.

\begin{figure}[!ht]
\includegraphics[width=0.483\textwidth]{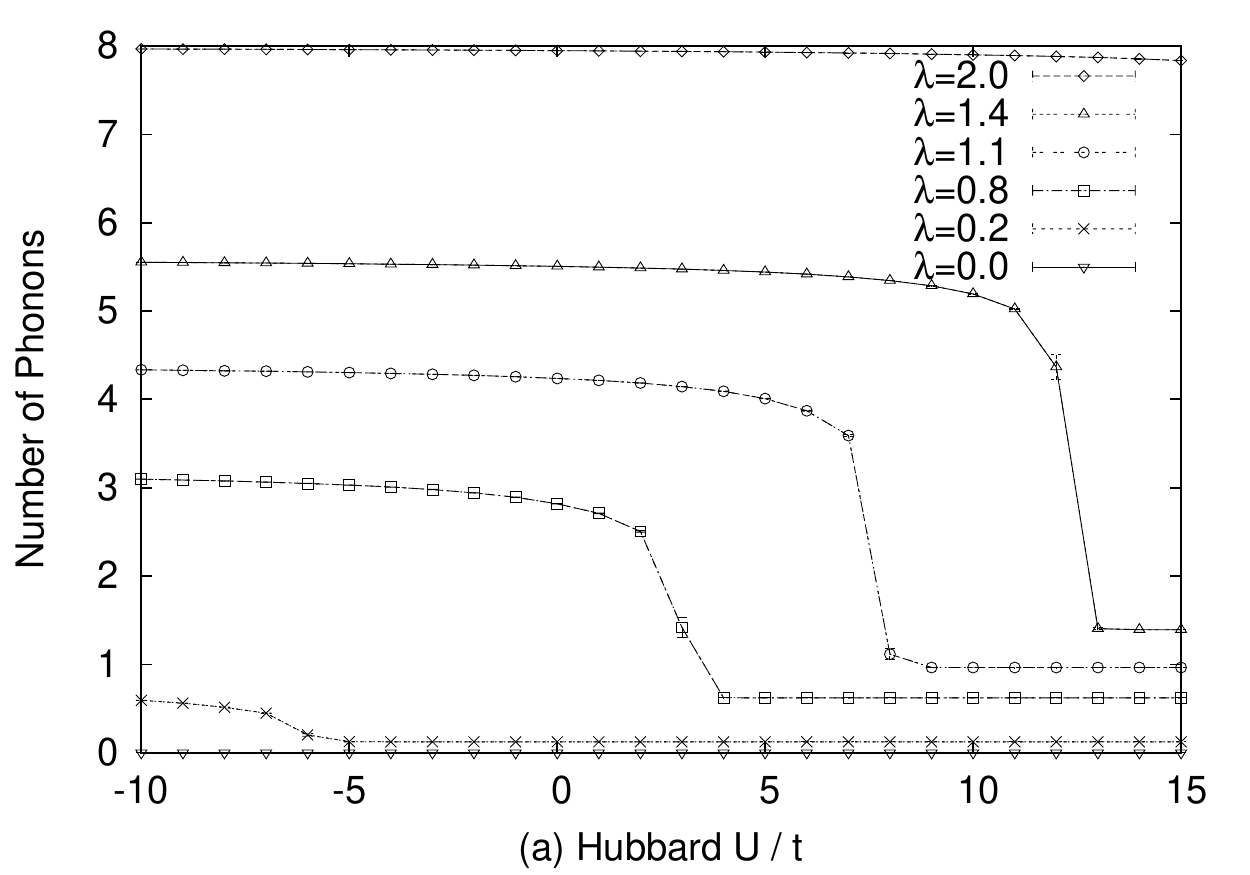}
\includegraphics[width=0.483\textwidth]{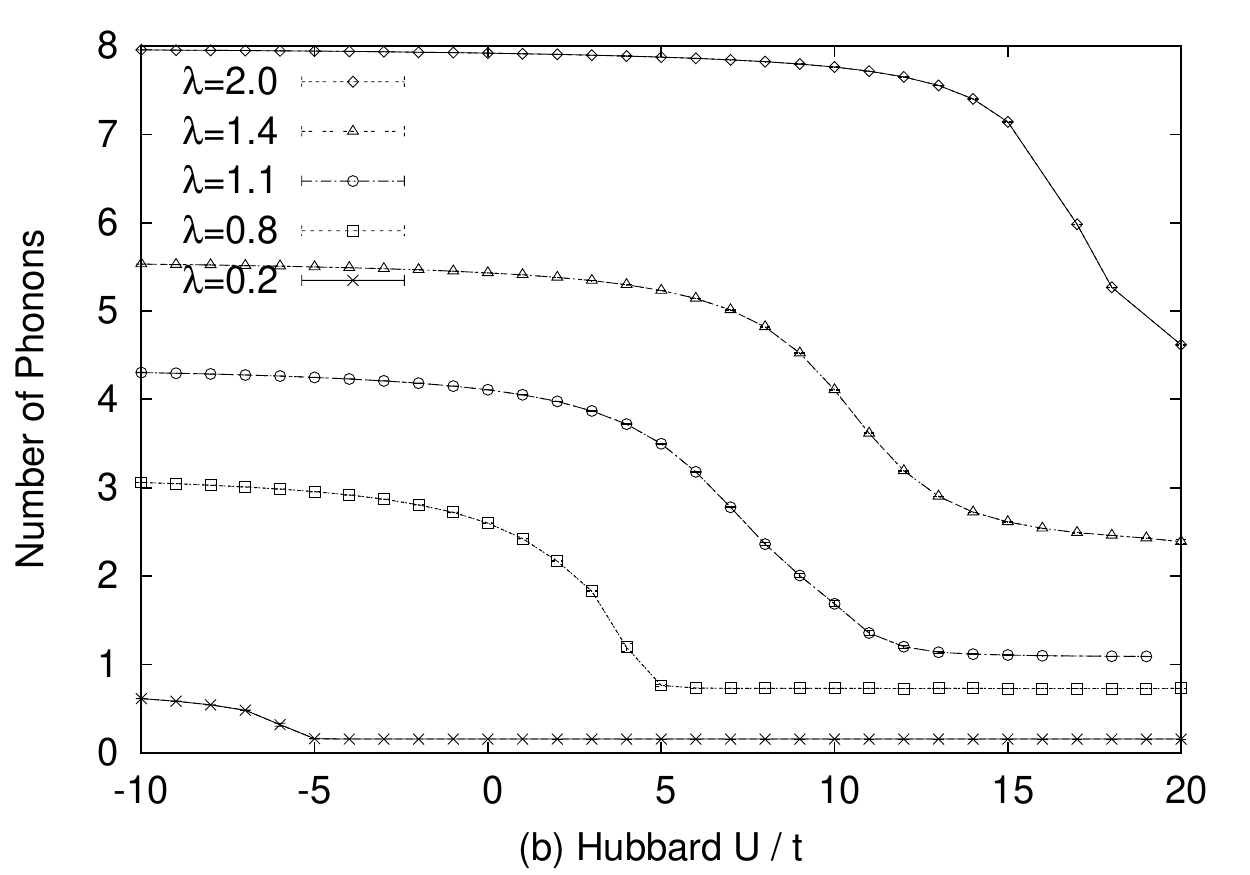} 
\includegraphics[width=0.483\textwidth]{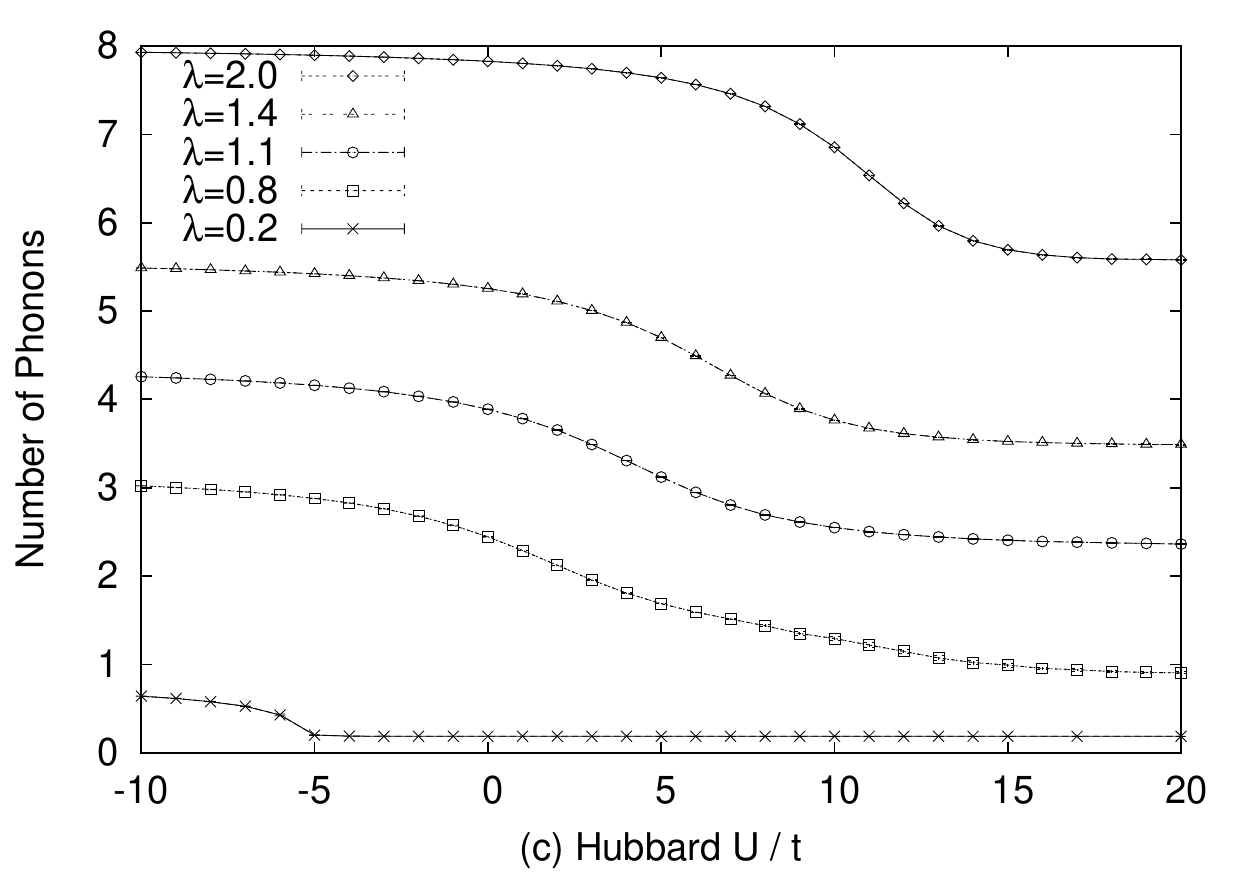}
\caption{Number of phonons associated with a singlet Hubbard--Holstein
bipolaron (a), bipolarons with nearest-neighbor
interaction strength of $\gamma = 0.25$ (panel b) and $\gamma = 0.5$
(c). Again, the range of crossover between on-site bipolaron
and unbound/inter-site bipolarons increases with $\gamma$.}
\label{Fig:CubicPhonons}
\label{fig:NumPhonons}
\end{figure}

Further evidence for binding of bipolarons can be found in the total
number of excited phonons shown in Figure \ref{Fig:CubicPhonons}. Here
we see that at large Hubbard $U$ the number of phonons in the system
does not change with respect to $U$, although it is non-zero, because
even unbound polarons have phonons associated with them. This can be
seen in panel (a) the Holstein case, as $\lambda$ increases from $\lambda = 0$ (where
there are no phonons) to higher phonon coupling where there are a
significant number of residual phonons at high $U$. The curve showing
the total number of phonons also levels out at large $U$ when there is
significant inter-site coupling and bipolarons large but bound, 
which occurs because there is vanishing on-site
component of the wave function in the presence of sufficient $U$ and
therefore the system is unchanged as $U$ is varied.

On lowering $U$ in Fig. \ref{fig:NumPhonons}(a), 
we see that the number of phonons associated with the
bipolaron rapidly increases as the on-site bipolaron forms. This is
due to the rapid crossover from unbound or inter-site pairs to on-site
pairs. As $U$ decreases further, the number of phonons tends again to
a set value dictated by the electron-phonon coupling constant and phonon frequency,
which also does not depend on $U$. This occurs at sufficient negative
$U$ where the bipolaron is forced into an on-site configuration. Plots
for $\gamma = 0.25$ and $0.5$ are shown in Figs. \ref{Fig:CubicEnergy}(b)
 and \ref{fig:NumPhonons}(c)  for the extended Holstein case. Again,
an increase in nearest neighbor attractive potential visibly smooths
out transition from bound pairs at low $U$ and the unbound or off-site
pairs seen at large $U$. The crossover from bound to unbound is
stretched over a larger range of $U$ consistent with the similar
observation in the total energy. With increased nearest-neighbor
attraction the number of associated phonons decreases less
dramatically as on-site pairs form, presumably because an increased
number of phonons are associated with the inter-site pairs.

\begin{figure}[!ht]
\includegraphics[width=0.483\textwidth]{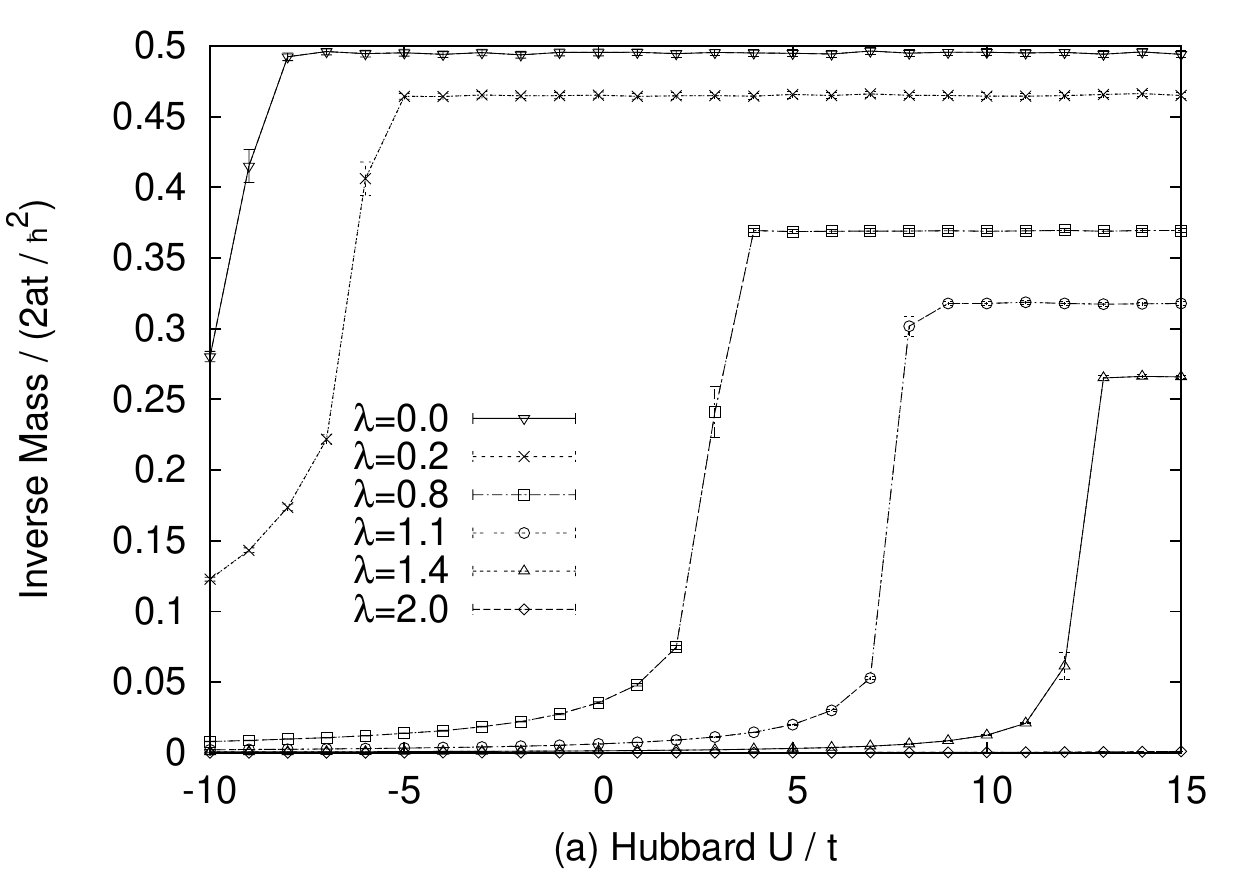}
\includegraphics[width=0.483\textwidth]{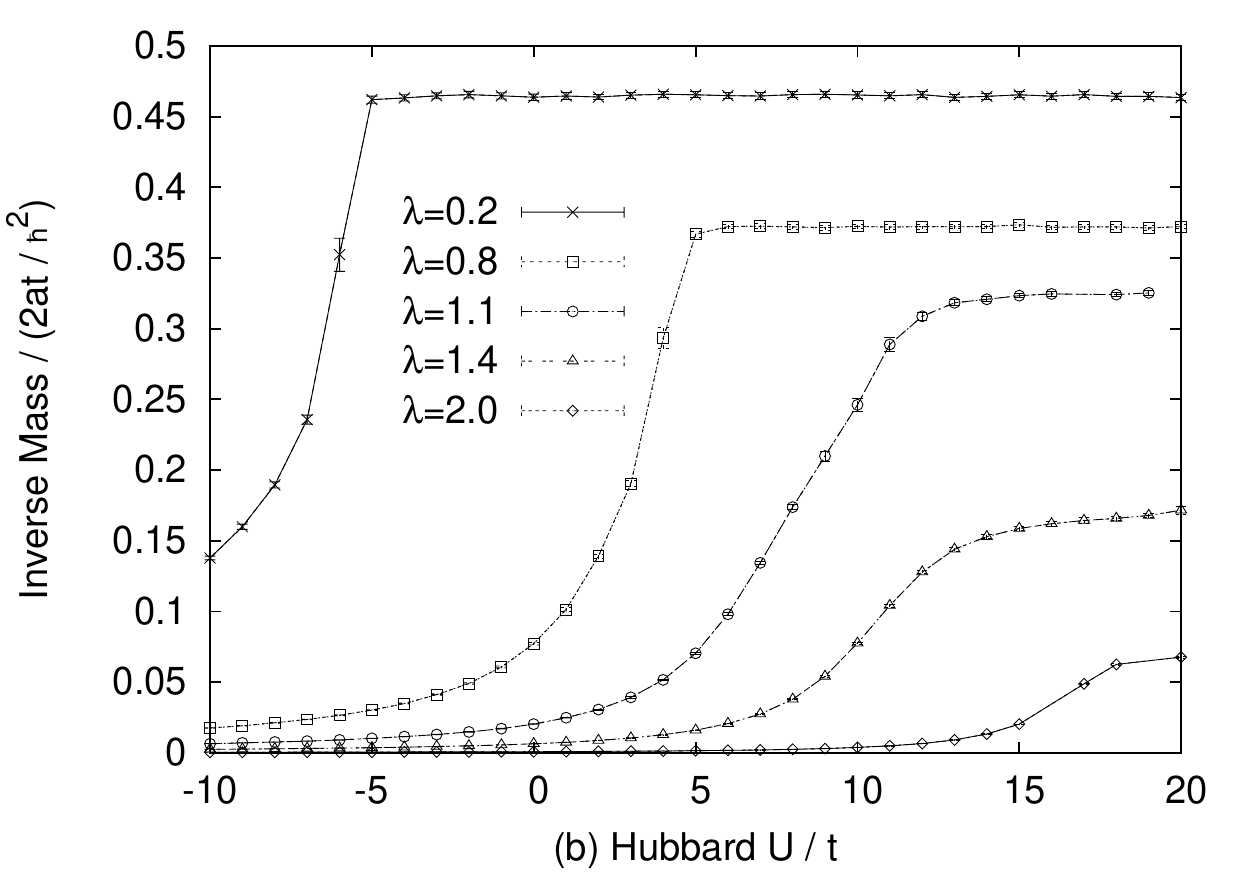} 
\includegraphics[width=0.483\textwidth]{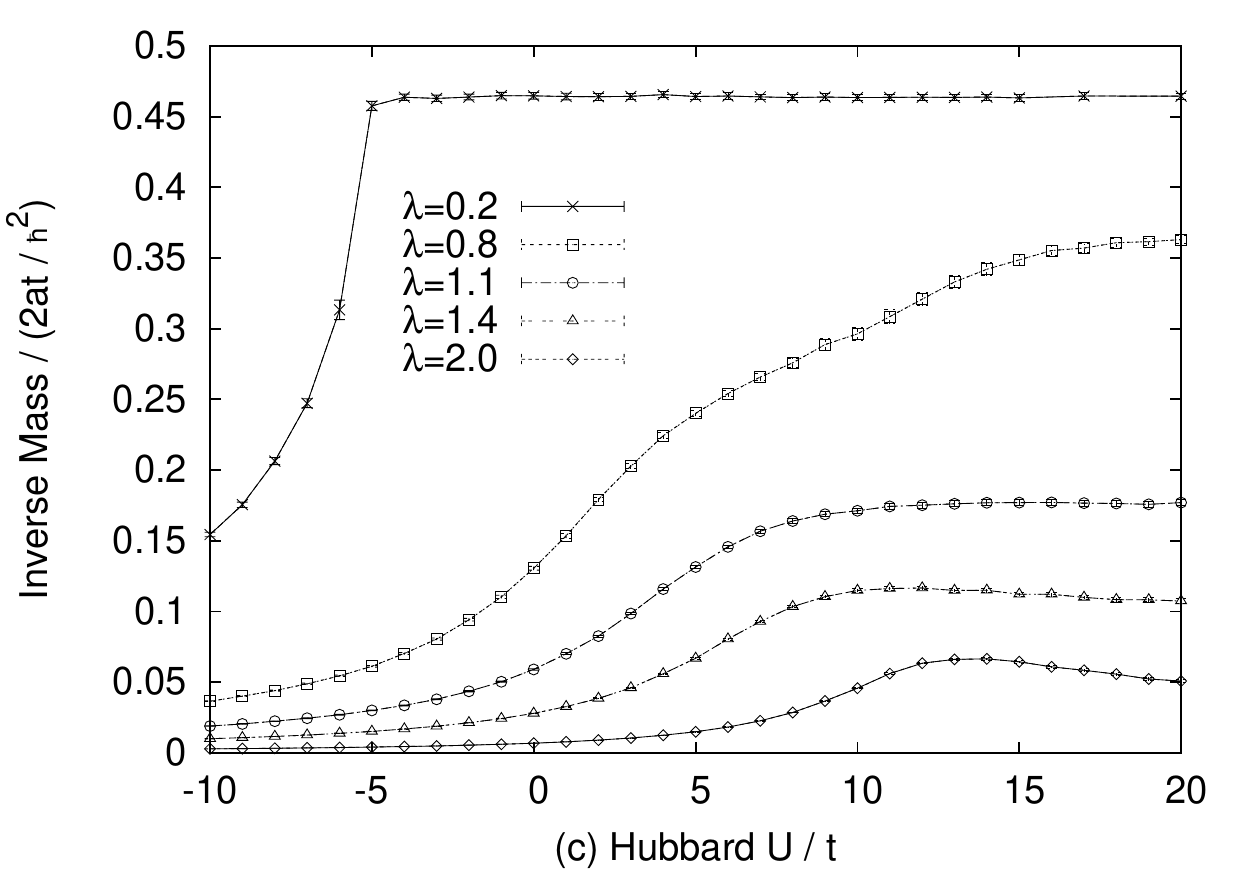}
\caption{Inverse mass of the singlet bipolaron with nearest-neighbor 
interaction for interaction strength of $\gamma=0$ (a), $\gamma = 0.25$ (b) and $\gamma
= 0.5$ (c). For large coupling constant $\lambda \gtrsim 1.1$ and
$\gamma = 0.5$, we see that the inverse mass has a maximum before
decreasing and then leveling off. The reduction in mass is a
version of the superlight small bipolaron behavior,
and is achieved when on-site and inter-site interactions become
comparable. Light mobile bipolarons may form a BEC with significant transition temperature.}
\label{fig:InverseMass}
\end{figure}

It is of particular interest to examine the change in singlet
bipolaron inverse mass as $U$ and $\lambda$ are varied, as this can be
related to the BEC transition temperature. Figure
\ref{fig:InverseMass} shows that the mass is near constant at large
$U$. Bound on-site pairs have a high mass (low inverse mass). The
inverse mass rapidly decreases at the point of binding. We see that
the transition from bipolaron to polaron starts at lower $U$, with
increased nearest neighbor interaction [Figs \ref{fig:InverseMass}(b) and \ref{fig:InverseMass}(c)]. The mass
decreases more slowly with increased Hubbard $U$ for a higher
interaction in accordance with the slow change in associated
phonons. With high coupling constant $\lambda \gtrsim 1.1$ and $\gamma
= 0.5$ we see that the inverse mass has a maximum before decreasing
and then leveling off, showing that the effective mass has a minimum
value at intermediate $U$. This phenomenon is not clearly visible in
the total energy. The reduction in mass is a version of the superlight
small bipolaron behavior
\cite{hague2007a,hague2007b,hague2009a,hague2010a} and is achieved
when on-site and inter-site interactions become comparable in size so
that bipolarons can hop by contracting and expanding through
degenerate on-site to inter-site pairs without energy penalty. This
behavior is significant because small mobile bipolarons could form a
Bose--Einstein condensate with significant transition temperature. We
will discuss this possibility in the next section.

\begin{figure}[!ht]
\includegraphics[width=0.483\textwidth]{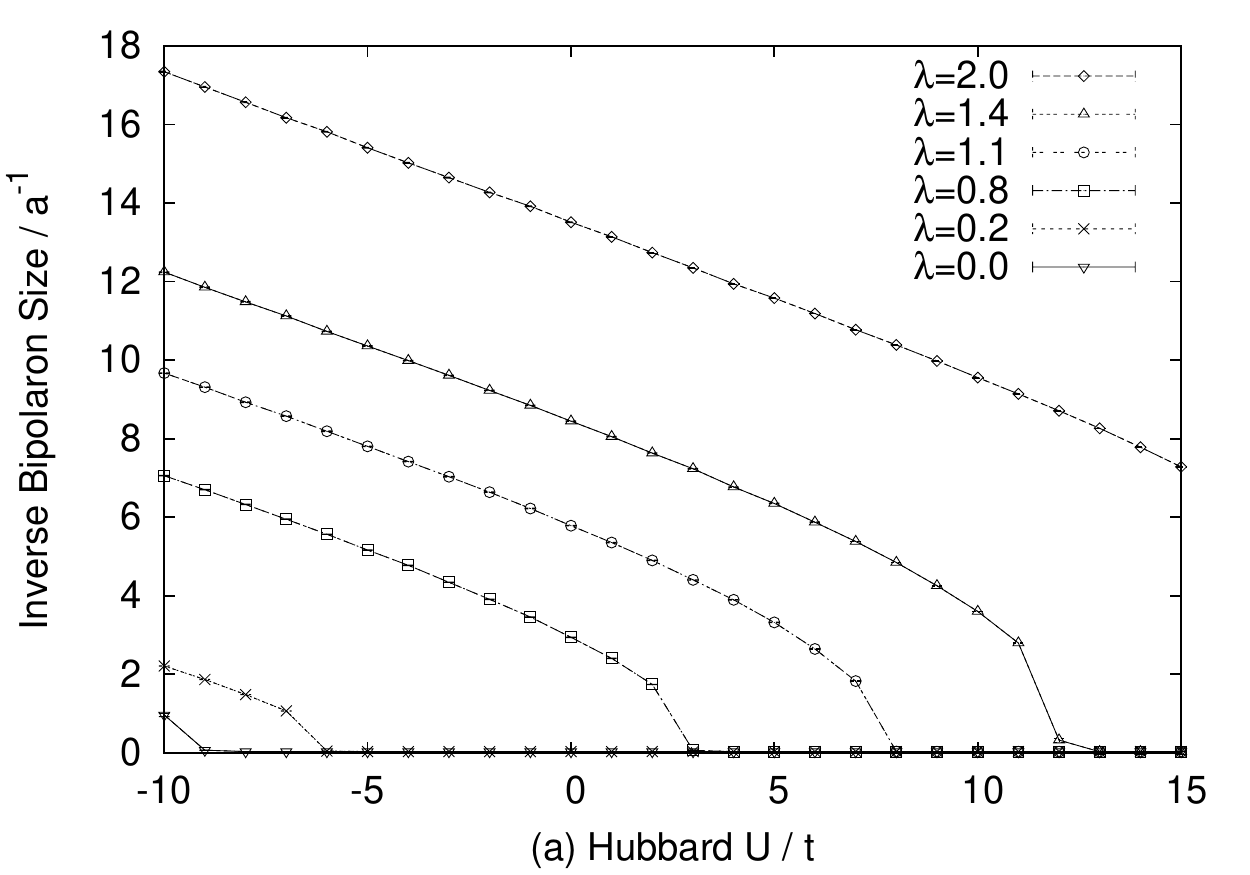}
\includegraphics[width=0.483\textwidth]{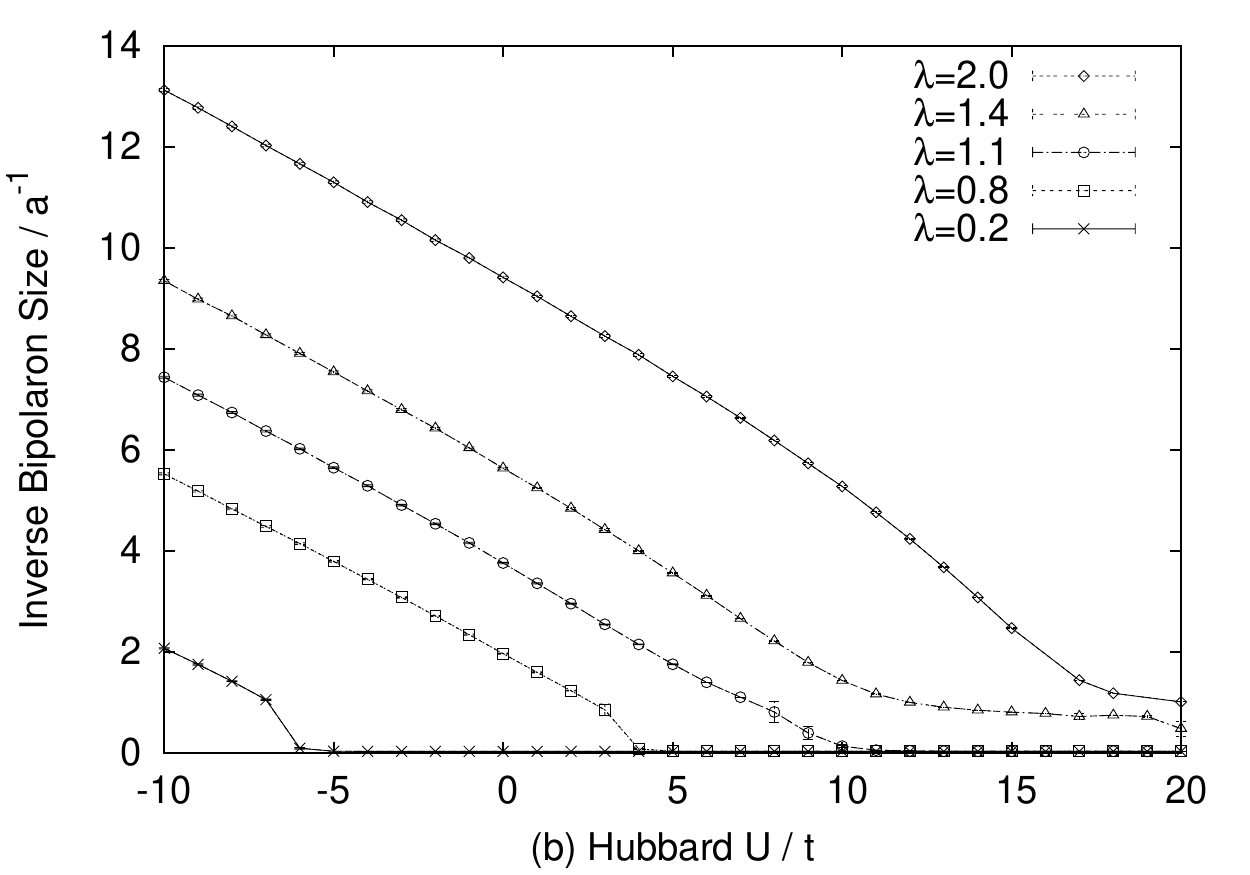} 
\includegraphics[width=0.483\textwidth]{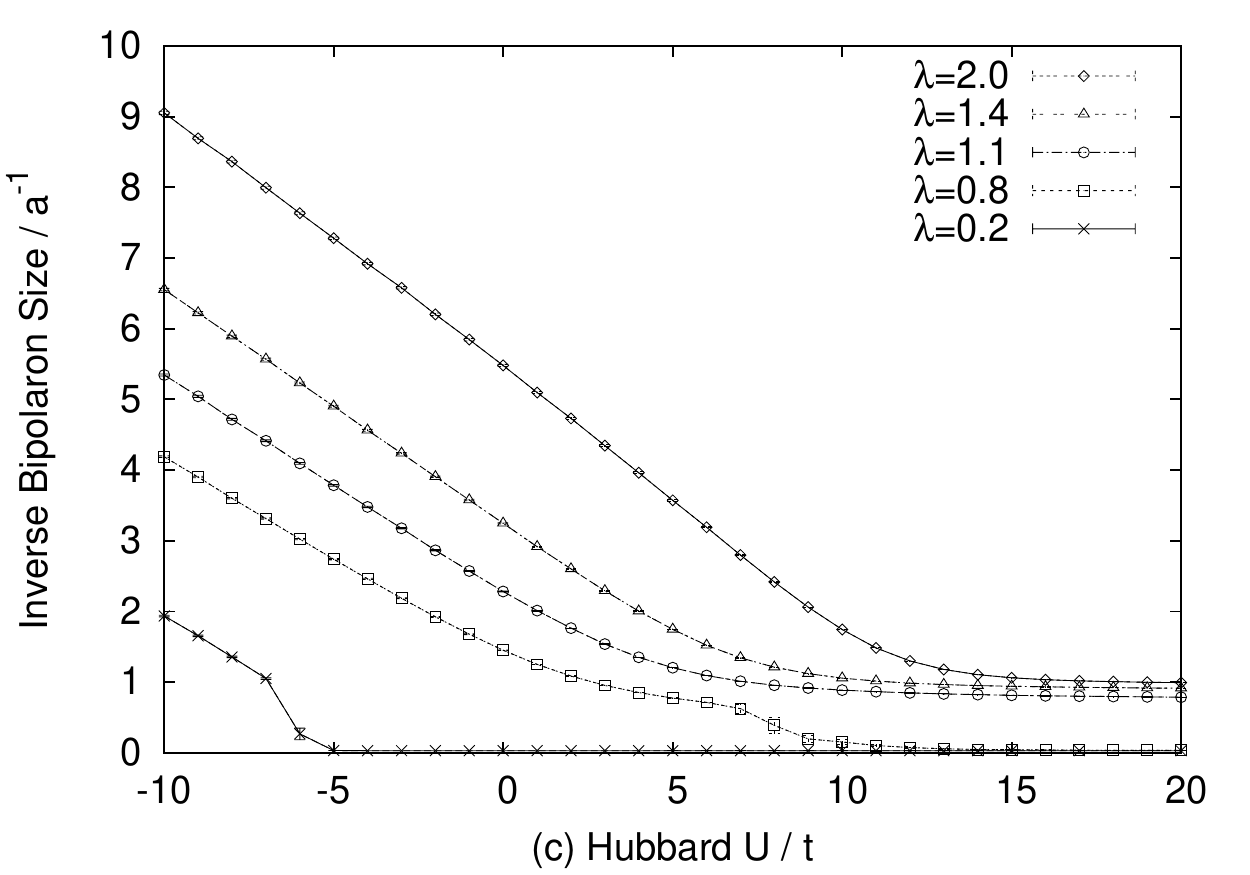}
\caption{Singlet bipolaron inverse size with nearest-neighbor interaction 
strength of $\gamma = 0$ (a), $\gamma = 0.25$ (b) and $\gamma = 0.5$ (c). At $\lambda=0.8$
and $\gamma=0.5$, an inflection in the bipolaron size shows the
transition of on-site bound pairs through inter-site pairing before the
bipolaron completely unbinds on increasing $U$. This is the precursor
of the superlight bipolaron behavior found at larger $\lambda$.}
\label{fig:InverseSize}
\end{figure}

Wavefunctions of individual pairs may not overlap if bipolarons are to
be well defined, so the bipolaron size limits the maximum density of
particles in a bipolaronic material. Figure \ref{fig:InverseSize}
plots the inverse average singlet bipolaron size against the Hubbard
$U$. Fig. \ref{fig:InverseSize}(a) shows the inverse bipolaron size for the local Holstein
interaction. As expected for large $U$, bipolaron pairs unbind and the
bipolaron size becomes infinite. With increasing coupling constant
$\lambda$ the average bipolaron size becomes smaller (inverse size
plotted) as the attractive phonon mediated interactions overcome the
repulsive Hubbard $U$.

In Figs. \ref{fig:InverseSize}(b) and \ref{fig:InverseSize}(c), where inter-site interaction is turned on,
qualitatively different behavior of the bipolaron size can be
seen. For weak electron-phonon coupling $\lambda \lesssim 1$ we see
that inverse bipolaron size tends to zero at high Hubbard $U$. The
value of $U$ required for unbinding increases with $\gamma$. With
large inter-site interaction of $\gamma=0.5$, we see that bipolaron
unbinding does not occur at high $U$ for coupling constant $\lambda
\gtrsim 1.1$ (within the range investigated), instead tending towards
a bipolaron size on the order of a lattice constant. At $\lambda=0.8$
and $\gamma=0.5$, an inflection in the bipolaron size shows the
transition of on-site bound pairs through off-site pairing before the
bipolaron completely unbinds on increasing $U$. This is the precursor
of the superlight bipolaron behavior found at larger $\lambda$

\begin{figure}[!ht]
\includegraphics[width=0.483\textwidth]{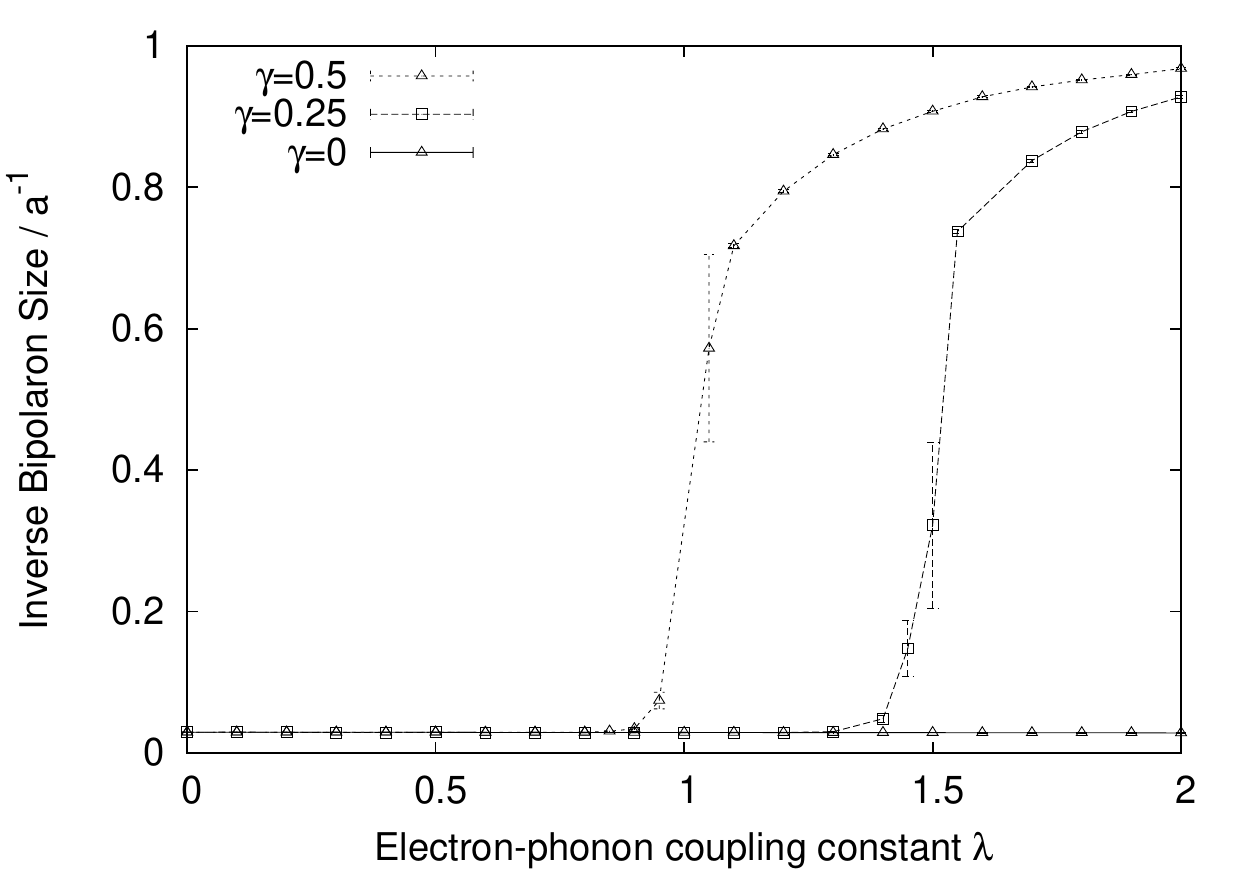}
\caption{Singlet bipolaron inverse size with increasing electron
  phonon coupling for infinite Hubbard $U$, showing the transition
  from unbound to bound states. Nearest neighbor strength $\gamma = 0, 
  0.25$ and $0.5$. Inter-site bound pairs are
  shown seen to exist at large $\lambda$ when nearest neighbor
  interactions are switched on.}
\label{fig:InfU_sim}
\end{figure}

We conclude this section by examining
the infinite Hubbard $U$ case. Figure \ref{fig:InfU_sim} displays
inverse bipolaron size against increasing electron-phonon coupling
constant with nearest neighbor interaction strength $\gamma = 0, 
  0.25$ and $0.5$, showing the crossover from
unbound states to bound inter-site pairs. It is seen that for
$\gamma=0.25$ and $\gamma=0.5$ the system is never fully unbound above
$\lambda\simeq1.5$ and $\lambda\simeq1$ respectively. Both have a
sharp decrease of bipolaron size that tends to a value equal to one
lattice spacing.  Inter-site bound pairs are therefore shown to exist
at large $U$ and $\lambda$ when nearest neighbor interactions are
switched on.

\section{Bose--Einstein condensation}
\label{Sec:super}

Evidence from the quantum Monte Carlo simulations presented in the
previous section shows that 3D bipolarons can be simultaneously small
and light in the region of the parameter space where perturbation
theory breaks down. In this section, we examine if bosonic charge
carriers of this type could form a Bose--Einstein condensate (BEC) with
a significant transition temperature.

The BEC transition temperature can be calculated using the expression,
\begin{eqnarray}
k_{\rm B}T_{\rm BEC} = \frac{3.31 \hbar^2}{m^{**}}\left(\frac{n_b}{a^3}\right)^{2/3},
\end{eqnarray}
where $n_b$ is the number of bosons per site, $m^{**}$ is the
effective boson mass and $a$ is the lattice constant (here we take $a$
to be 4.2\AA, consistent with the bismuthates). An upper bound on the
number of bosons per lattice site can be established by utilizing the
bipolaron size $R$, $n_b/a^3 \approx 1/R^{\prime3}$, where 
$R^\prime$ is the effective radius of the bipolaron. From this we get the following relation,

\begin{eqnarray}
T_{\rm BEC} = \frac{3.31 \hbar^2}{k_{\rm B}m^{**}R^{\prime2}}.
\end{eqnarray}

\begin{figure}[!ht]
\includegraphics[width=0.483\textwidth]{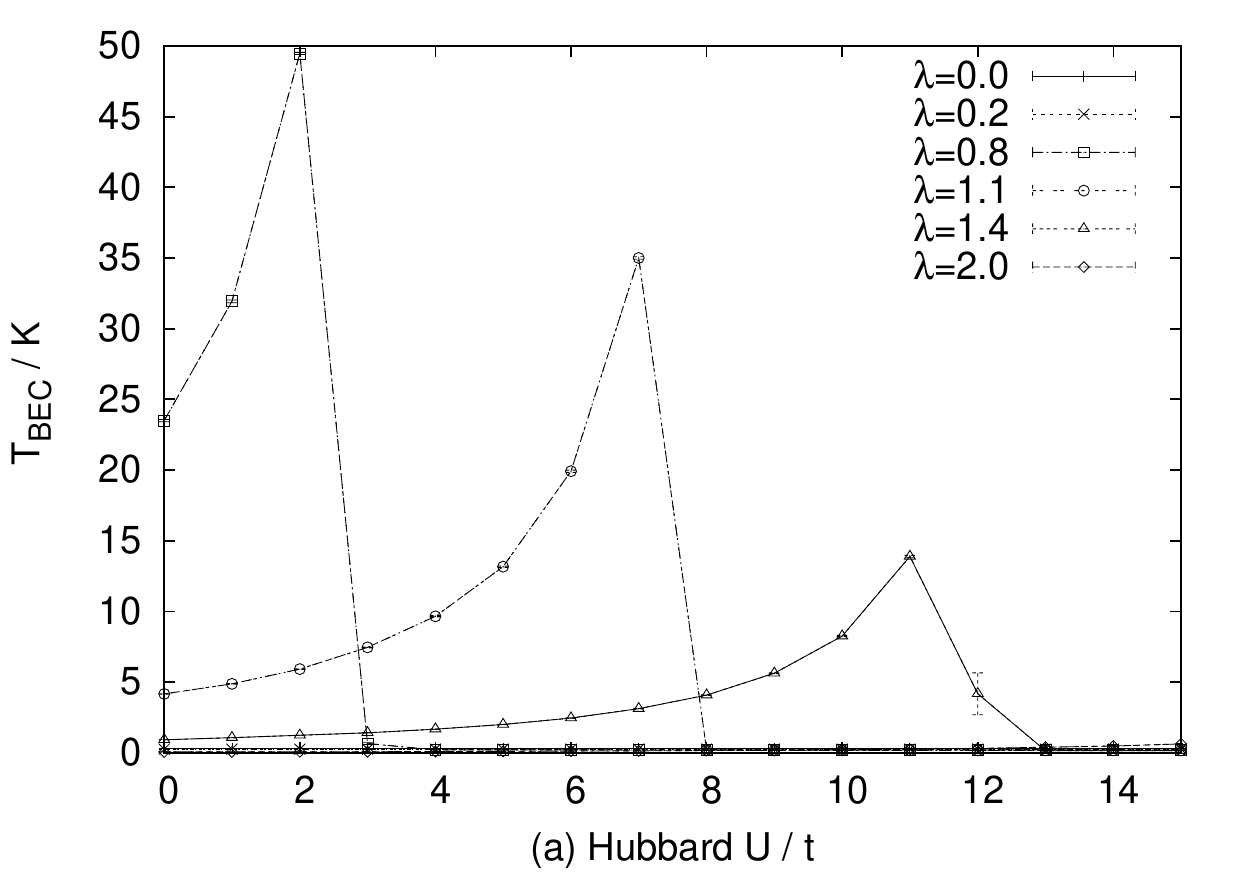}
\includegraphics[width=0.483\textwidth]{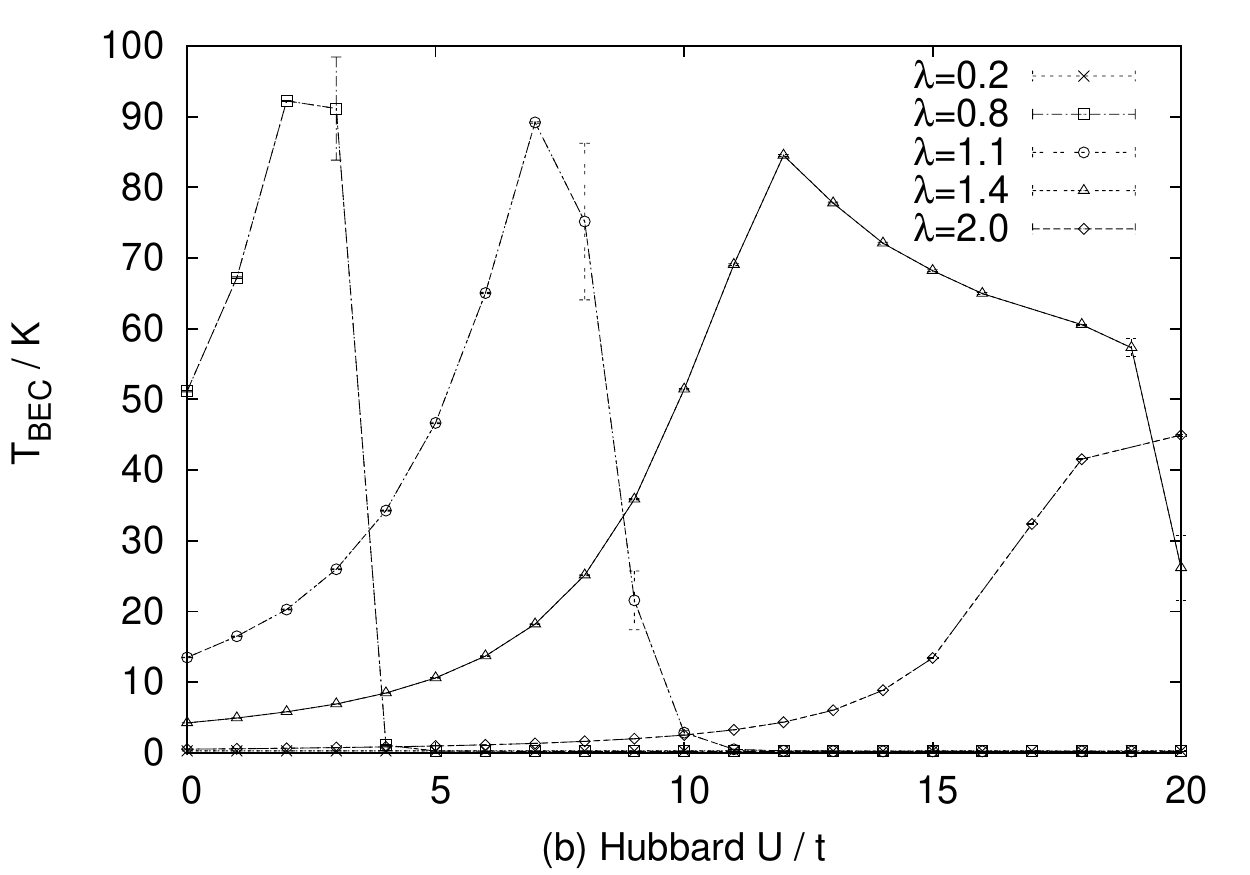} 
\includegraphics[width=0.483\textwidth]{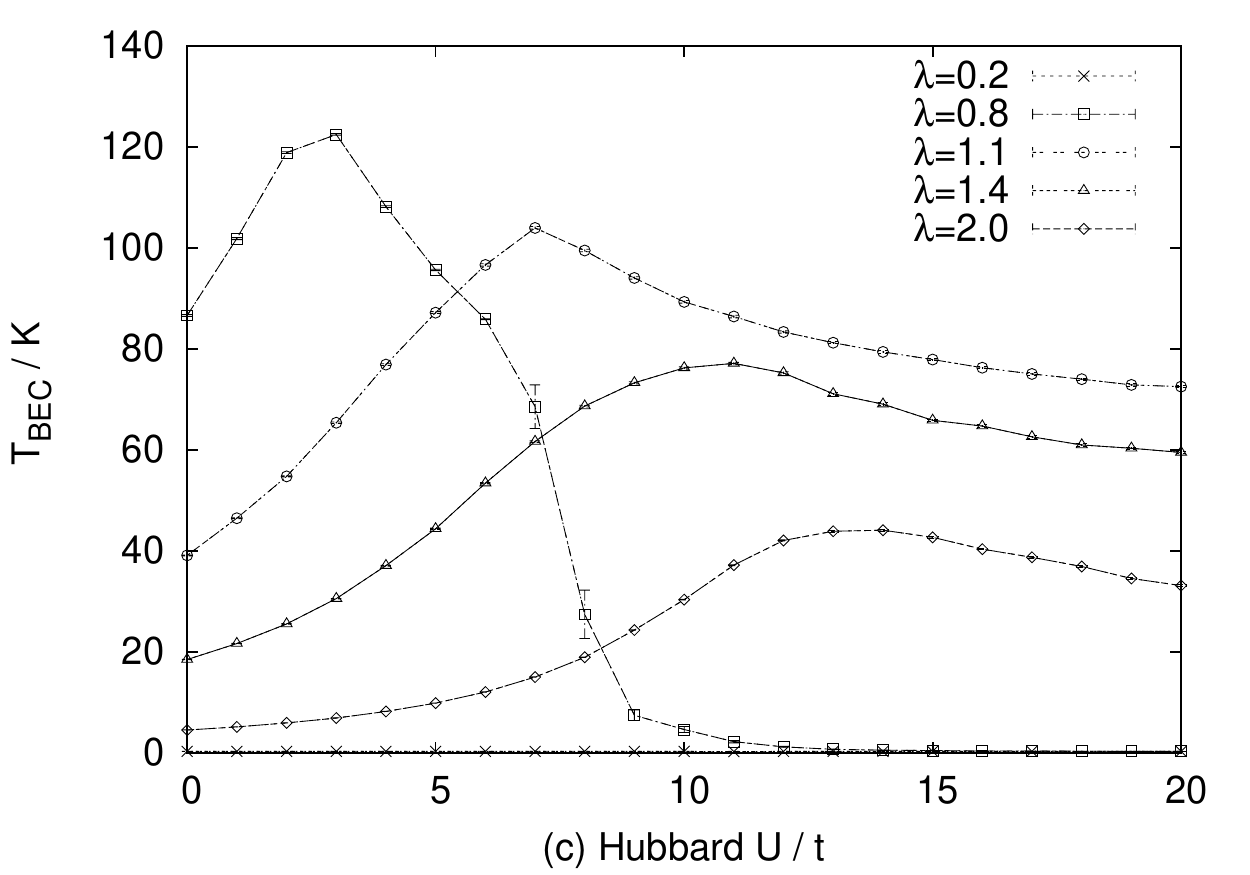}
\caption{BEC transition temperature with nearest-neighbor interaction
  strength of $\gamma = 0$ (panel a), $\gamma = 0.25$ (panel b) and
  $\gamma = 0.5$ (panel c).  The divergence in $T_{\rm BEC}$ for the
  Holstein case will be suppressed because the bipolaron is only weakly bound. On the other hand, high transition temperatures of 20-30K are seen with
  nearest neighbor interaction strength $\gamma=0.5$ and medium
  electron-phonon coupling constant where bipolarons are well bound.}
\label{fig:Tc}
\end{figure}

Figure \ref{fig:Tc} is a plot of transition temperatures with (a)
Holstein and inter-site interaction strengths (b) $\gamma=0.25$ and
(c) $\gamma=0.5$. We only plot positive Hubbard $U$ in this section,
since negative $U$ would be unphysical. The effective radius 
$R^\prime$ is taken to be 5$R$ (a bipolaron separation is 5 bipolaron radii in 
each direction) corresponding to a bipolaron wavefunction overlap of less than 1$\%$. 
As bipolaron densities increase so that bipolarons overlap, interaction corrections are
expected to reduce transition temperatures.

Several sharp peaks in $T_{\rm BEC}$ are seen in Fig. \ref{fig:Tc}(a)
for Holstein electron-phonon coupling constants of $\lambda=0.8, 1.1$
and $1.4$ (for weaker $\lambda$, the peaks are at unphysical negative
$U$ values). These peaks reach values temperatures of $\sim50K$. The
regions of high $T_{\rm BEC}$ are unstable to small variation in $U$
and bipolarons are only weakly bound, so thermal fluctuations will
suppress transition temperatures by breaking up the bipolaron. For
weak electron-phonon coupling, bipolarons only bind at unphysical
negative $U$ values, and no BEC is formed for positive $U$. It is
interesting to note that although bipolarons are formed through
electron-phonon coupling, there is a wide region of the parameter
space where increasing the Hubbard $U$ raises the transition
temperature.

BEC transition temperatures are shown for nearest neighbor interaction
$\gamma=0.25$ in Figure \ref{fig:Tc}(b). Similar to the Holstein case,
$T_{\rm BEC}$ has a peak at low Hubbard $U$ for intermediate
$\lambda$. The peak width increases with electron-phonon coupling, but
the maximum in the transition temperature decreases slightly. A tail
appears at large $U$ for coupling constant $\lambda=1.4$, where the
bipolaron becomes bound between sites and properties depend only
weakly on $U$. QMC simulations are essential here, since the
regions of high transition temperature (where both the electron-phonon
coupling and Hubbard $U$ are intermediate) can not be accessed using
perturbative techniques.

Finally Fig. \ref{fig:Tc}(c) plots BEC transition temperatures for
nearest-neighbor interaction strength $\gamma=0.5$. The strong
inter-site coupling completely eradicates the sharp peaks in transition
temperature associated with the Holstein interaction, replacing them
with broad continuous curves with increased $T_{\rm BEC}$. Bipolarons
with medium to high $\lambda$ form stable nearest-neighbor pairs with
low effective masses, leading to superconducting states that have
significant transition temperatures over wide range of $U$. Bipolarons
formed from large electron-phonon coupling have larger effective
masses, leading to significantly lower condensation temperatures. The
most interesting point here is that for medium-sized coupling
constants, bipolaron effective masses are still small when bipolarons
are bound into small inter-site pairs, resulting in high condensation
temperatures of $90-120K$. Note that the use of $R^\prime$ leads to
an approximation on the possible $T_{\rm BEC}$, lower
transition temperatures are estimated if the upper bound on the
distance between bipolarons is larger (before inter-boson interactions need to be
taken into account), and interactions between bipolarons typically lower transition temperatures. Again, it is interesting to
note that the BEC transition temperature can increase as $U$
increases, even though the mechanism for binding pairs is phonon
mediated. Since bipolarons are bound at very large $U$ and $\lambda$
(as shown in Figure \ref{fig:InverseSize}) no breakdown of the BEC is
seen for very large repulsive Coulomb interactions, and the transition
temperature remains significant. This is important because many oxide
materials with large electron-phonon interactions also have large $U$.

\begin{figure}[!ht]
\includegraphics[width=0.483\textwidth]{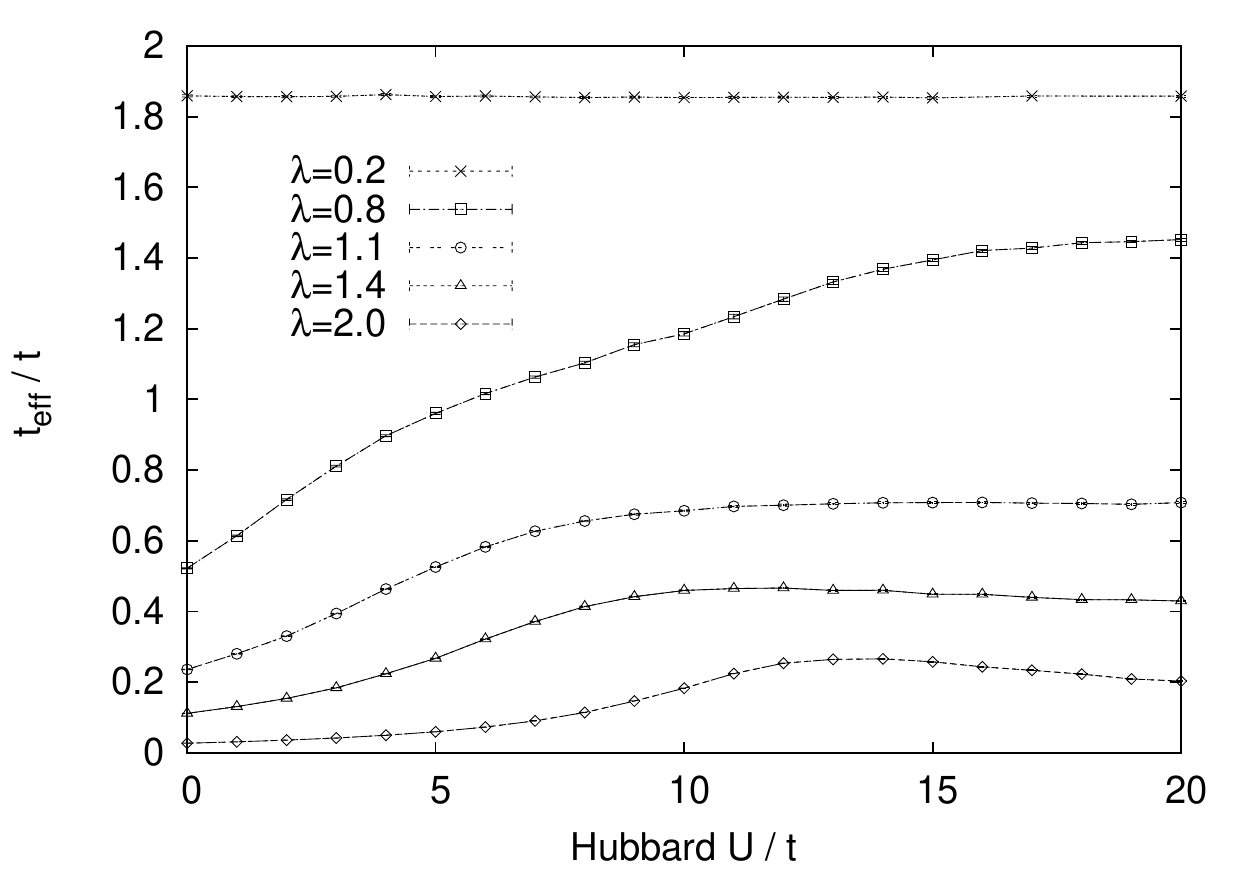}
\caption{Effective Hopping energies with nearest-neighbor interaction strength of $\gamma = 0.5$. Medium electron-phonon coupling $\lambda$ here are shown to be mobile with effective hopping energies in the region of the bare hopping energy.}
\label{fig:Hop}
\end{figure}

To probe the sensitivity of bipolarons to impurities, we calculate the
effective hopping for the bipolarons,
\begin{eqnarray}
t^\prime = \frac{\hbar^2}{2a^2m^{**}}.
\end{eqnarray}
Figure \ref{fig:Hop} shows the effective bipolaron hopping energy when
$\gamma=0.5$.  For electron-phonon coupling constants $\lambda=1.1$
and $\lambda=1.4$, bipolarons have an effective hopping $t_{\rm
  eff}\approx 0.7t$ and $t_{\rm eff}\approx 0.4t$ respectively. As we
have previously shown \cite{hague2009a}, local impurities with energy
$\Delta=-4t_{\rm eff}$ are needed to pin polarons to
impurities. Therefore, for any reasonable impurity size, bipolarons are mobile.

\section{Summary and conclusions}

We have investigated the formation of bipolarons and their subsequent
Bose--Einstein condensation on a three dimensional cubic lattice. A
quantum Monte Carlo code was employed to investigate regimes of
intermediate electron-phonon coupling and Coulomb repulsion, and was
validated using analytic calculations in the large phonon frequency
limit. Away from the regions where perturbation theories are valid,
the effective mass and bipolaron radius are consistent with light
small bipolarons.

A consequence of the 3D lattice is that binding of bipolarons is
difficult for weak $\lambda$. Small and mobile bipolarons form for
intermediate $\lambda$ and inter-site coupling $\gamma$ when the
energies of on-site and inter-site pairs become similar. By analyzing
the exact numerical results, we have shown that bipolaron condensation
temperatures (leading to superconductivity) could be up to $90-120$ K
for realistic bipolaron densities. Another consequence of 3D is that
it is more difficult to bind bipolarons to impurities. Impurity
energies of around $-4t_{\rm eff}$ are required to localize particles
in 3D. Therefore, the light bipolaron states are stable against
attractive impurity levels with energies of up to $\sim t$ as the
effective hopping has a similar magnitude to the bare electron hopping
energy.  We conclude that stable bipolaron superconductors that are
insensitive to changes in Hubbard $U$ could form in three-dimensional
oxides with inter-site electron-phonon interactions of intermediate
magnitude (that is, electron-phonon interactions with a moderate
momentum dependence). Moreover, with sufficiently large inter-site
electron-phonon coupling, superconductivity could be stable at very
large values of $U$, demonstrating that Coulomb repulsion is no
barrier to bipolaronic superconductivity in 3D.

\section*{Acknowledgments}
We are pleased to thank Andrey Umerski, Sasha Alexandrov and John
Samson for useful discussions. JPH would like to acknowledge EPSRC
grant No. EP/H015655/1.

\bibliography{Bibliography}

\appendix

\section{High phonon frequency and the UV model}
\label{sec:HighFrequency}
The bipolaron properties can be analytically approximated in the high
phonon frequency (anti-adiabatic) limit $( \hbar \omega \gg W)$ by
using the result of the Lang--Firsov transformation, since if the
phonon frequency is very large there are no real phonons. Up to a linear shift in energy the resulting $U-V$ model is shown as:
\begin{multline}
\tilde{H} = - \sum_{nn^\prime\sigma} t'_{nn^\prime} c^\dagger_{n\sigma} 
c_{n^\prime\sigma} + U'\sum_n c^\dagger_{n\uparrow} c_{n\uparrow} c^\dagger_{n\downarrow} 
c_{n\downarrow} \\ + \sideset{}{^\prime}\sum_{nn^\prime} \sum_{\sigma\sigma^\prime} V'_{nn^\prime} 
c^\dagger_{n\sigma} c_{n\sigma}c^\dagger_{n^\prime\sigma'} c_{n^\prime\sigma'} \label{eqn:UVham}
\end{multline}
The primed sum over $V'$ in the final part of Equation
(\ref{eqn:UVham}) ignores the self-interaction term. The interaction
terms in this Hamiltonian for on-site interaction and nearest neighbor
interactions are $U' = U - 2W\lambda$ and $V'_{nn^\prime} = 2W\lambda
\chi_0(n,n^\prime) / \chi_0(0,0)$ respectively.

Taking the two particle Schr\"odinger equation,
\begin{multline}
[E - \epsilon(\kvec_1)-\epsilon(\kvec_2)]\chi(\kvec_1,\kvec_2) \\
= U' \sum_{\qvec} \chi(\qvec,\kvec_1 + \kvec_2 - \qvec) \\ - 
V'\sum_{\lvec} e^{-i\kvec_1\cdot\lvec}\sum_{\qvec} \chi(\qvec,\kvec_1 + 
\kvec_2 - \qvec)e^{i\qvec\cdot\lvec}
\label{eqn:fourier}
\end{multline}
\small\begin{eqnarray}
\epsilon(\kvec) = -t' \sum_{\lvec} e^{-i\kvec\cdot\lvec} = 
-2t'(\cos{\kvec_x}+\cos{\kvec_y}+\cos{\kvec_z})
\end{eqnarray}\normalsize
where $\lvec= \{(\pm1,0,0),(0,\pm1,0),(0,0,\pm1)\}$ assuming that
the lattice constant $a=1$ (this will be assumed throughout). To simplify the problem
we introduce a set of momentum dependent values, $\Delta(\Kvec)$,
\begin{gather}
\Delta_{(0,0,0)}(\Kvec) \equiv \sum_{\qvec} \chi(\qvec,\kvec_1 + \kvec_2 - \qvec) \\
\Delta_{\lvec}(\Kvec) \equiv \sum_{\qvec} \chi(\qvec,\kvec_1 + \kvec_2 - \qvec) e^{i\qvec\cdot\lvec}
\end{gather}
and then substitute them into the Schr\"odinger equation
(\ref{eqn:fourier}). Rearranging the resulting equation, we obtain an
expression for $\Phi$ in terms of $\Kvec$ and $\qvec$, where
$\Kvec = \kvec_1 + \kvec_2$:
\small
\begin{eqnarray}
\chi(\kvec_1,\kvec_2) &=& \frac{U'\Delta_{(0,0,0)}(\Kvec) - 
V' \sum_{\lvec} \Delta_{\lvec}(\Kvec) e^{-i\kvec_1\cdot\lvec}}
{E - \epsilon(\kvec_1)-\epsilon(\kvec_2)} \\
\chi(\qvec,\Kvec-\qvec) &=& \frac{U'\Delta_{(0,0,0)}(\Kvec) - 
V' \sum_{\lvec} \Delta_{\lvec}(\Kvec) e^{-i\qvec\cdot\lvec}}
{E - \epsilon(\qvec)-\epsilon(\Kvec-\qvec)}
\label{eqn:Phi}
\end{eqnarray}
\normalsize
Expanding the expression for $\chi(\qvec,\Kvec-\qvec)$ from Equation (\ref{eqn:Phi}) 
into a matrix format leads to the following relation:
\small\begin{widetext}\begin{gather}
\left( \begin{array}{ccccccc}
L_0 - \frac{1}{U'} & L_{-x} & L_{x} & L_{-y} & L_{y} & L_{-z} & L_{z} \\
L_{x} & L_0 + \frac{1}{V'} & L_{2x} & L_{x-y} & L_{x+y} & L_{x-z} & L_{x+z} \\
L_{-x} & L_{-2x} & L_0 + \frac{1}{V'} & L_{-x-y} & L_{-x+y} & L_{-x-z} & L_{-x+z} \\
L_{y} & L_{y-x} & L_{y+x} & L_0 + \frac{1}{V'} & L_{2y} & L_{y-z} & L_{y+z} \\
L_{-y} & L_{-y-x} & L_{-y+x} & L_{-2y} & L_0 + \frac{1}{V'} & L_{-y-z} & L_{-y+z} \\
L_{z} & L_{z-x} & L_{z+x} & L_{z-y} & L_{z+y} & L_0 + \frac{1}{V'} & L_{2z} \\
L_{-z} & L_{-z-x} & L_{-z+x} & L_{-z-y} & L_{-z+y} & L_{-2z} & L_0 + \frac{1}{V'}
\end{array} \right)
\left( \begin{array}{c}
U'\Delta_{(0,0,0)} \\
-V'\Delta_{x} \\
-V'\Delta_{-x} \\
-V'\Delta_{y} \\
-V'\Delta_{-y} \\
-V'\Delta_{z} \\
-V'\Delta_{-z} \\
\end{array} \right)
=0 \label{eqn:LMatrix}
\end{gather}\end{widetext}\normalsize
\begin{eqnarray}
L_p = L_{q_p}(E,\Kvec) = \sum_{\qvec} \frac{e^{iq_p}}{E - \epsilon(\qvec)-\epsilon(\Kvec-\qvec)}
\label{LS}
\end{eqnarray}
At the $\Gamma$ point, $\Kvec=(0,0,0)$, the additional symmetry
simplifies analysis of Eq. (\ref{eqn:LMatrix}), since
Eq. (\ref{LS}) can be reduced to the following components, 
$L_{\pm x}   = L_{\pm y}   = L_{\pm z} \equiv L_1$, 
$L_{\pm 2x} = L_{\pm 2y} = L_{\pm 2z} \equiv L_2$, and
$L_{\pm x \pm y} = L_{\pm y \pm z} = L_{\pm z \pm x} \equiv L_3$. 

To diagonalize the problem, a new basis of
states with $s$,$p$ and $d$ wave symmetry is introduced;
\small\begin{equation}\centering\begin{split}
\Delta_0 &= \Delta_{(0,0,0)} \\
\Delta_s = \frac{1}{\sqrt6} (\Delta_{x}+\Delta_{-x}&+\Delta_{y}+\Delta_{-y}+\Delta_{z}+\Delta_{-z})\\
\Delta_{p_1} =& \frac{1}{\sqrt2} (\Delta_{x}-\Delta_{-x})\\
\Delta_{p_2} =& \frac{1}{\sqrt2} (\Delta_{y}-\Delta_{-y})\\
\Delta_{p_3} =& \frac{1}{\sqrt2} (\Delta_{z}-\Delta_{-z})\\
\Delta_{d_1} = \frac{1}{\sqrt4} (\Delta_{x}&+\Delta_{-x}-\Delta_{y}-\Delta_{-y}) \\
\Delta_{d_2} = \frac{1}{\sqrt{12}} (\Delta_{x}+\Delta_{-x}&+\Delta_{y}+\Delta_{-y}-2\Delta_{z}-2\Delta_{-z})
\end{split}\end{equation}\normalsize
where the $s$ and $d$ states are symmetric and $p$ states antisymmetric on inversion through the origin.

Applying the new basis of $\Delta$'s to the matrix equation leads to a simple
block diagonal matrix consisting of two $s$ states
($\Delta_0$ and $\Delta_s$), three $p$ and two $d$
states.
\footnotesize\begin{eqnarray}
\left( \begin{array}{ccccccc}
L_0 - \frac{1}{U'} & 6L_{1} & 0 & 0 & 0 & 0 & 0 \\
L_{1} & L_{S} + \frac{1}{V'} & 0 & 0 & 0 & 0 & 0 \\
0 & 0 & \lambda_p & 0 & 0 & 0 & 0 \\
0 & 0 & 0 & \lambda_p & 0 & 0 & 0 \\
0 & 0 & 0 & 0 & \lambda_p & 0 & 0 \\
0 & 0 & 0 & 0 & 0 & \lambda_d & 0 \\
0 & 0 & 0 & 0 & 0 & 0 & \lambda_d
\end{array} \right)
\left( \begin{array}{c}
U'\Delta_{0} \\
-V'\Delta_{s} \\
-V'\Delta_{p_1} \\
-V'\Delta_{p_2} \\
-V'\Delta_{p_3} \\
-V'\Delta_{d_1} \\
-V'\Delta_{d_2} \\
\end{array} \right)
\label{NewMatrix}
\end{eqnarray}\normalsize
where,

\begin{equation}\centering\begin{split}
L_0 =& \sum_{\qvec} \frac{1}{E-2\epsilon(\qvec)} \label{L0} \\
L_{S} =& \sum_{\qvec} \frac{2\cos{q_x}(\cos{q_x}+\cos{q_y}+\cos{q_z})}{E-2\epsilon(\qvec)} \\
\lambda_p =& L_0 + \frac{1}{V'} - L_2 \\
\lambda_d =& L_0 + \frac{1}{V'} + L_2 - 2L_3
\end{split}\end{equation}
The $p$ and $d$ states are diagonalized in the new basis and their
energies can be calculated directly from,

\begin{equation}\begin{split}
\lambda_p  = 0 \\
\lambda_d  = 0
\label{eqn:PandDstates}
\end{split}\end{equation}
The ground state singlet states can be computed 
from solution of the $2 \times 2$ matrix in the 
top left hand corner of Eq.
(\ref{NewMatrix}):
\begin{eqnarray}
\left( \begin{array}{cc}
L_0 - \frac{1}{U'} & 6L_{1} \\
L_{1} & L_{S} + \frac{1}{V'}
\end{array} \right)
\left( \begin{array}{c}
U'\Delta_{0} \\
-V'\Delta_{s} \\
\end{array} \right)
= 0,
\end{eqnarray}
which can be used to calculate energies of the $s$ state by taking the determinant,
\begin{eqnarray}
\left( L_0 - \frac{1}{U'} \right) \left(L_{S} + \frac{1}{V'} \right) - 6L_{1}^2 = 0.
\label{eqn:det}
\end{eqnarray}
Rearranging Eq. (\ref{eqn:det}),
\begin{eqnarray}
L_0 = \frac{V'(6U'L_{1}^2 + L_{S}) + 1}{U'(V'L_{S} + 1)},
\label{eqn:NewL0}
\end{eqnarray}
a binary search can be used to
determine values of $E$ for various values of $U'$ and $V'$.

\begin{figure}[ht]
\centering
\includegraphics[width=0.483\textwidth]{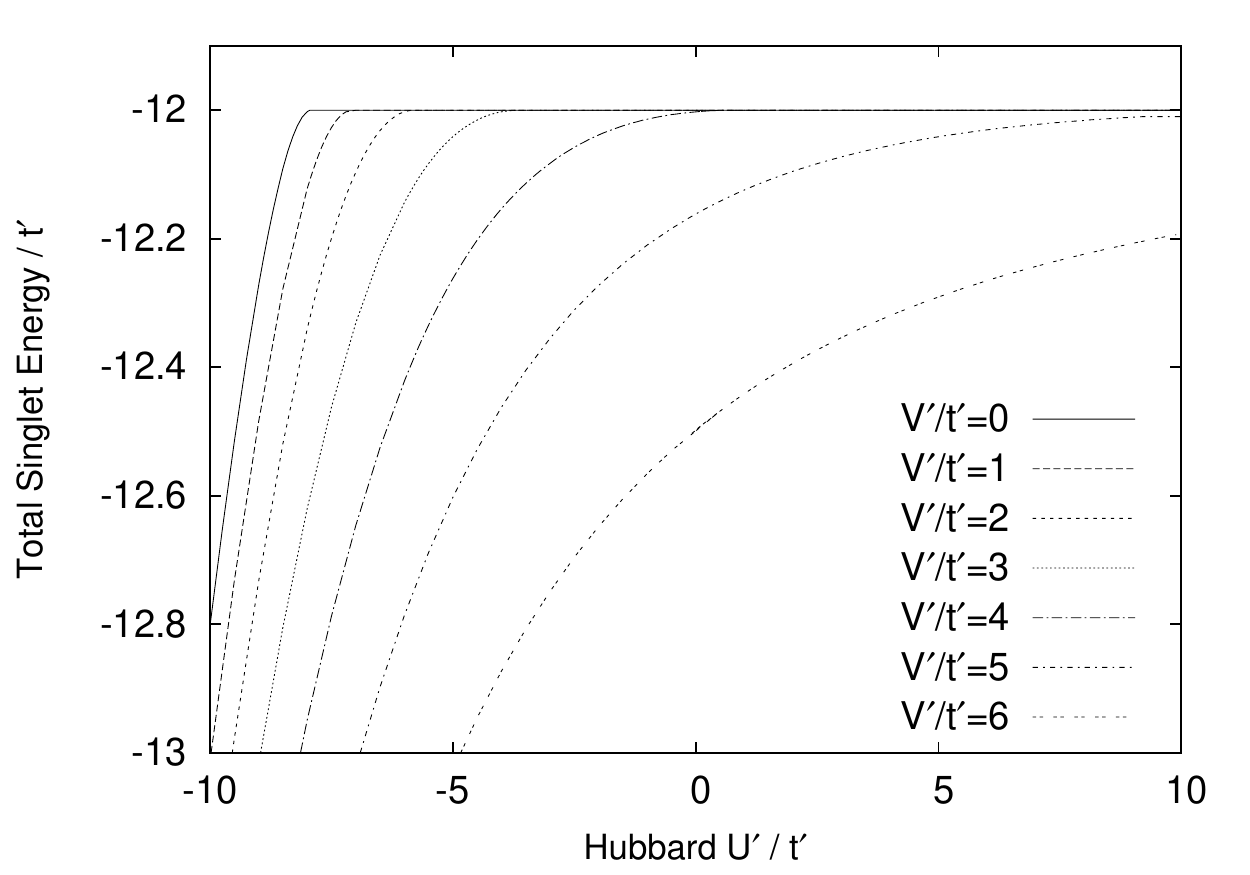}
\caption{Ground state bipolaron singlet energy computed for a $UV$ model on a cubic
lattice. A key result here is that it takes a large negative Hubbard
$U$ to bind on-site bipolarons in contrast to 1D and 2D systems. At
low $V'$, the bipolaron binds for finite $U$, but for large
$V'$, the unbinding occurs at very large (infinite) $U$ since
the inter-site bipolaron is stable.}
\label{fig:Analytical Cubic UV}
\end{figure}

The solutions found by this method are shown in Fig.
\ref{fig:Analytical Cubic UV}. Typically, there is a smooth transition
from bound states (diagonal lines) to unbound states (horizontal
lines). For $V=0$ the graph depicts a diagonal line with constant
gradient representing a bound on-site pair that at about $U'/t'\simeq -8$
levels off to a horizontal line representing unbound polarons in the
lattice. With increasing nearest-neighbor potential $V'$ the transition
from bound to unbound states spans a larger range of Hubbard $U'$
(curved line), and is related to the presence of inter-site pairs in the
lattice. At large enough $V'$ there are no unbound states.

Even if there is no inter-site repulsion, $V'/t'=0$, strong negative
Hubbard values $U'/t' =-7.915$ are required to bind the bipolaron in
contrast to 1D and 2D lattices where you need $U'/t' = 0$
\cite{hague2009a,hague2010a}. This is due to the additional degrees
of freedom in the cubic structure, where electrons are not confined in
any direction. For a nearest neighbor attraction of $V'/t'=4$ Figure \ref{fig:Analytical Cubic UV}
shows that binding occurs around $U'=0$ (actual crossing value $V'/t'=3.875$), whereas on the square lattice
the presence of $V'/t'$ leads to off-site pairing at strong positive
$U'$. The lack of confinement in 3D mean that bipolaron on-site pairing
is not guaranteed even with high coupling constants. Applying a small
nearest-neighbor potential in 2D has a much bigger effect on the binding
than in 3D due to the confinement \cite{hague2010a}.

\begin{figure}[ht]
\centering
\includegraphics[width=0.483\textwidth]{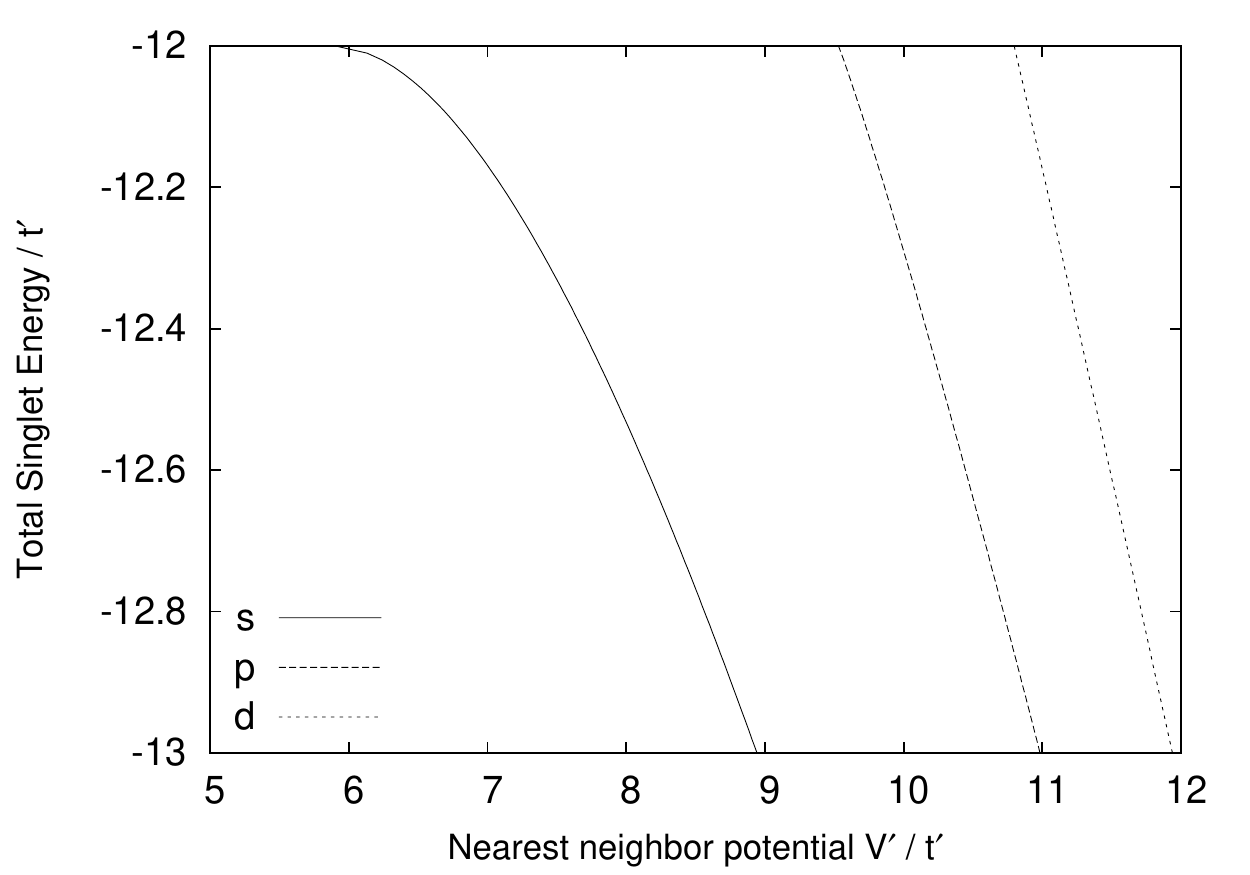}
\caption{Exact bipolaron energy computed for a $U-V$ model on a cubic
lattice. Here we examine the energy at infinite $U'$ to observe if there
is a critical $V'$ that ensures binding in the 3D lattice for $s$, $p$
and $d$ states.}
\label{fig:Analytical Cubic Inf U}
\end{figure}

To understand the qualitative difference in bipolaron behavior as $V'$
is changed, we evaluated the energy of the pair at infinite $U$. In
Fig. \ref{fig:Analytical Cubic Inf U} we plot the total energy (from
Equation (\ref{eqn:NewL0})) as a function of nearest neighbor
potential $V'$, showing the binding of the $s$ state as $V'$ is
increased. We also show both $p$ and $d$ states for completeness. For
the $s$ state, the binding crossover begins at a nearest neighbor
interaction strength of $V'/t'=5.875$. After this point the energy
curves sharply to an approximate $E\propto V'$ for high $V'$. The
energies of the $p$ and $d$ states do not level off at $E=-12t'$,
since they are excited states.

\end{document}